\pdfoutput=1
\documentclass[12 pt]{article}
\usepackage{titling}
\usepackage{blindtext}
\usepackage{graphicx}
\usepackage{hyperref}
\usepackage[font=small,labelfont=bf]{caption}
\usepackage[left=1in,right=1in,top=1.1in,bottom=1.1in]{geometry}
\newcommand{\minus}{\scalebox{0.75}[1.0]{$-$}}
\usepackage[utf8]{inputenc}
\usepackage{amsmath}
\usepackage{subcaption}
\usepackage[affil-it]{authblk}
\numberwithin{equation}{section}
\usepackage[nodisplayskipstretch]{setspace}
\DeclareUnicodeCharacter{2212}{-}
\usepackage{mathtools}
\usepackage{booktabs}
\newcommand{\Lagr}{\mathcal{L}}
\usepackage[style=trad-unsrt, citestyle=numeric-comp]{biblatex}
\usepackage[table]{xcolor}
\usepackage{breakcites}

\hypersetup{
  colorlinks   = true, 
  urlcolor     = black, 
  linkcolor    = blue, 
  citecolor   = blue
}

\let\oldabstract\abstract
\let\oldendabstract\endabstract
\makeatletter
\renewenvironment{abstract}
{%
               {\endlist}%
\oldabstract}
{\oldendabstract}
\makeatother
    
\title{\Large{\bf{Variational Autoencoders for Jet Simulation}}}
\date{\vspace{-3.5ex}}
\author{Kosei Dohi \\ \href{mailto:21kdohi@tenafly.k12.nj.us}{21kdohi@tenafly.k12.nj.us}}
\affil{Tenafly High School \\ Tenafly, NJ 07670 USA}

\begin{document}
\begin{titlingpage}
    \maketitle
    \begin{abstract}
    We introduce a novel variational autoencoder (VAE) architecture that can generate realistic and diverse high energy physics events. The model we propose utilizes several techniques from VAE literature in order to simulate high fidelity jet images. In addition to demonstrating the model’s ability to produce high fidelity jet images through various assessments, we also demonstrate its ability to control the events it generates from the latent space. This can be potentially useful for other tasks such as jet tagging, where we can test how well jet taggers can classify signal from background for events generated by the VAE. We test this idea by seeing the signal efficiency vs background rejection for different types of jet images produced by our model. We compare our VAE with generative adversarial networks (GAN) in several ways, most notably in speed. The architecture we propose is ultimately a fast, stable, and easy-to-train deep generative model that demonstrates the potential of VAEs in simulating high energy physics events.
    \end{abstract}
\end{titlingpage}
{
  \hypersetup{linkcolor=black}
  \tableofcontents
}

\noindent\hrulefill

\section{Introduction}

Deep generative models used in high energy physics simulation have shown promising performances due to their ability to produce data in record times~\cite{oliveira2017learning, sipio2019dijetgan, collaboration2019fast, paganini2017calogan, otten2019event, carrazza2019lund, paganini2017accelerating, alonsomonsalve2018imagebased, martinez2019particle, bellagente2019gan, butter2019gan, hashemi2019lhc, chekalina2018generative, Carminati_2018, musella2018fast, belayneh2019calorimetry, oliveira2017controlling, pawlowski2018reducing, erbin2018gans, erdmann2018precise, alanazi2020aibased, derkach2019cherenkov, refId0, zhou2018regressive, Vallecorsa_2018, erdmann2018generating, butter2020ganplifying}. Many of the calorimeter-based events that produce 2D images, such as jets and electromagnetic showers, utilize generative adversarial networks (GAN) to generate these events~\cite{cogan2014jetimages, komiske2016deep, oliveira2015jetimages}. Although GANs have demonstrated to excel in simulating diverse and accurate radiation patterns, there are various other models that can be explored that can achieve similar results and alleviate previous problems with GANs. One problem with GANs, for example, is that they can be difficult to train and produce classes of images that are too distinguishable from each other~\cite{oliveira2017learning}. An alternative model that can solve these potential problems is the variational autoencoder (VAE)~\cite{kingma2013autoencoding}. The VAE is an extension of the autoencoder (AE) and consists of the same principal components of the autoencoder -- the encoder, the latent space, and the decoder. In addition to these components, the VAE adds a sampling layer that allows the model to generate new data from the latent space. More detail can be found in section \ref{Variational Autoencoder}.

The application of GANs in various domains have highlighted its strong ability to produce crisp images~\cite{ledig2016photorealistic, wang2017highresolution}. This has ultimately made the GAN a popular choice for image generation. On the other hand, VAEs are known to produce noisier images due to the nature of its latent space and how images are generated from it. However, with recent progress in VAE research, VAEs have become comparable to GANs in producing highly realistic images for a wide variety of fields~\cite{razavi2019generating, vahdat2020nvae, yi2020cosmovae, pu2016variational, gulrajani2016pixelvae, cheng2020variational, Higgins2017betaVAELB}. A good example of VAEs specifically used in high energy physics is for generating electromagnetic and parton showers~\cite{Ghosh:2680531, buhmann2020getting}. Models such as the Introspective VAE~\cite{huang2018introvae} and the Vector-Quantised VAE~\cite{oord2017neural} have also shown to produce high resolution face images that are comparable to the state-of-the art GANs. Autoencoders and unsupervised/weakly supervised machine learning models, in general, have been applied to a wide variety of fields including in high energy physics~\cite{lim2018molecular, semeniuta2017hybrid, sinai2017variational, khoshaman2018quantum, voloshynovskiy2019information, bhalodia2019dpvaes, mackey2015fuzzy, komiske2019metric, dillon2020learning, dillon2019uncovering, dery2017weakly, metodiev2017classification, komiske2018learning, collins2018anomaly, collins2019extending, borisyak2019machine, cohen2017machine, komiske2018operational, metodiev2018topic, amram2020tag, cerri2018variational, farina2018searching, heimel2018qcd}. 

One of the key characteristics of the VAE that allows it to be a diverse generative model is its latent space. With its latent space, the VAE can control what kind of features it generates. A simple example with face images would be adding facial features, such as a smile, to the images being generated by the VAE. Following a similar idea, calorimeter deposits can be fed through a VAE to produce images with certain types of features such as the outer radiation of a jet.

In this paper, we propose a novel VAE architecture that can be used to simulate jet images. The model uses a variety of techniques from VAE/GAN research to help generate the sparse features of jet images. We test the model through certain qualitative and quantitative assessments and also observe the latent space to see what kind of features the VAE learns and how it can be used to simulate jets with diverse properties. 

\section{Methods} 

The methods section is organized as follows. We start by describing how vanilla autoencoders and variational autoencoders work. Afterward, we briefly describe the dataset we used to train the model. And finally, we talk about the architecture of the model we propose and its training process.

\subsection{Variational Autoencoder}\label{Variational Autoencoder}

To provide a better understanding of the variational autoencoder, we first provide a brief overview of the autoencoder. The autoencoder consists of three main components: the encoder, the decoder, and the latent space. The encoder takes input data, in our case, a dataset of images, and compresses it into the latent space using neural networks. The latent space is where all the information is stored and the goal is to optimize how data is scattered in this latent space. The decoder, on the other hand, is a mirror of the encoder because it tries to recreate the input data from the compressed latent space. 
\begin{figure}[h!]
  \centering
  \includegraphics[trim=24 14 100 30,clip, width=1\textwidth]{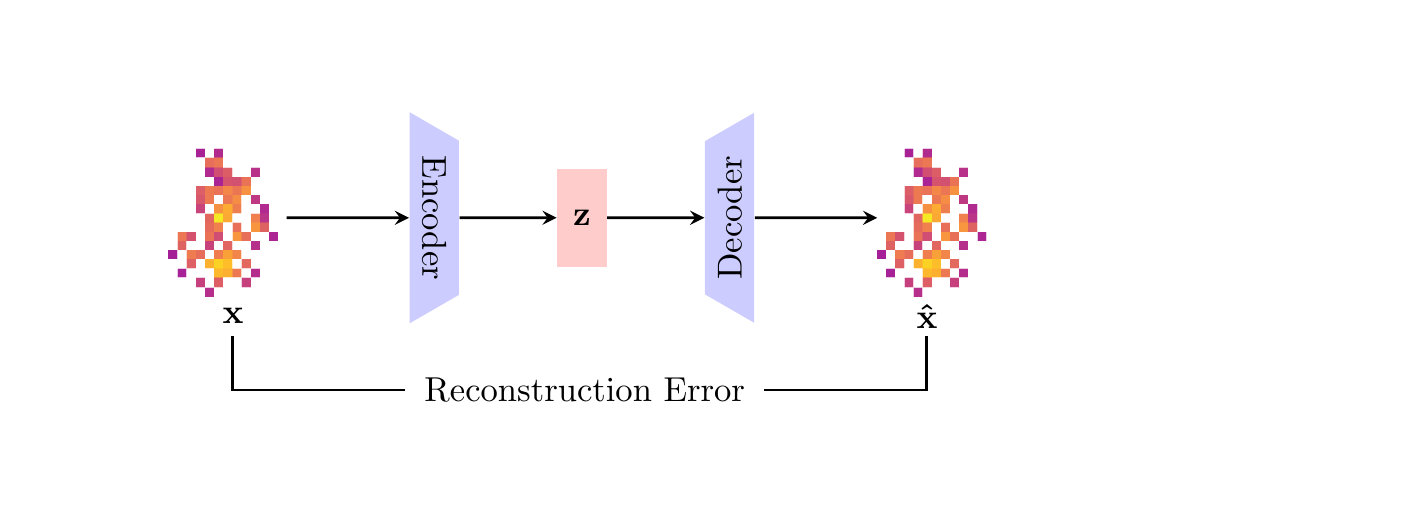}
  \caption{Simple schematic of the autoencoder. The input image, denoted by $X$,  is fed through the encoder and into a latent dimension $z$. The decoder then tries to reconstruct the input image from the latent space to get a predicted output image denoted by $\hat{X}$.}
\end{figure}

The goal of the autoencoder is to ultimately provide a dimensionality reduction algorithm that learns to encode data by optimizing a distance metric loss. The better the encoder learns to compress data, the better the reconstructed image will be. A common loss function used for the autoencoder is the mean squared error, which takes the squared difference between the input and output image. 

\begin{equation}
\frac{1}{n}\sum_{i=1}^{n}(\hat{X} - X)^2
\end{equation}

A variational autoencoder is a generative model that follows a similar structure to the vanilla autoencoder. The reason why a normal autoencoder isn't able to generate new data is because it encodes input data into discrete values in the latent space. This process only allows the autoencoder to ``memorize" the input data. To contrast this, the variational autoencoder encodes the input data into a sampling layer that makes the latent space continuous. This process allows the decoder of the variational autoencoder to generate new and realistic data using the learned features from the latent space. The latent space of the variational autoencoder can be explored because images with certain features in the latent space will be closer together than ones in another region.

\begin{figure}[h!]
  \centering
  \includegraphics[trim=35 20 110 27,clip, width=1\textwidth]{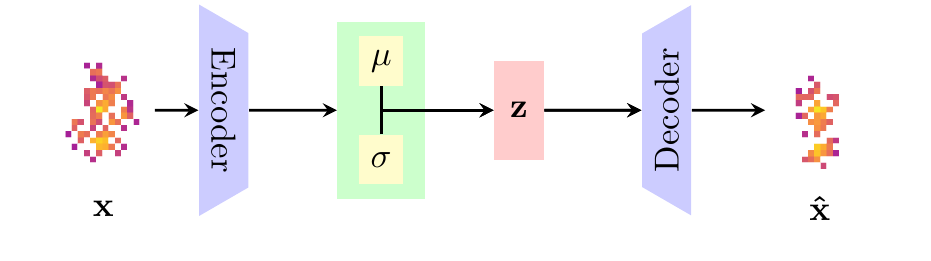}
  \caption{Simple schematic of the variational autoencoder. The input image is fed through the encoder and into a sampling layer, represented by the green box. The sampling layer maps the image into a Gaussian distribution with a certain mean ($\mu$) and standard deviation ($\sigma$). This allows the decoder to sample from a continuous latent space, or, in other words, enables it to generate images using features from input $X$.}
\end{figure}

The objective of the VAE can be mathematically described with the following equation:

\begin{equation}
{\Lagr(\theta, \phi; X)} = \overbracket{\underbrace{E[\log P_\theta(X|z)]}_\text{\clap{Reconstruction term}} - \underbrace{D_{KL}[Q_\phi(z|X)||P_\theta(z)]}_\text{\clap{KL Divergence}}}^{\mathclap{\text{Variational Lower Bound}}} \label{eq:1}
\end{equation}
Where the encoder system is denoted by $Q_\phi(z|X)$, the latent space is denoted by $z$, and the decoder system is denoted by $P_\theta(X|z)$. The Kullback-Divergence loss ($D_{KL}$), which is a loss function that helps ensure that the output of the encoder doesn't deviate from the posterior distribution $P(z)$, can be described with the following equation
\begin{equation}
D_{KL} = \frac{1}{2}\sum_k (\exp(\sigma) + \mu^2 - 1 - \log(\sigma))
\end{equation}
Where $\mu$ and $\sigma$ represents the mean and standard deviation of the Gaussian distribution used to generate data. The $E[\log P(X|z)]$ term in the variational autoencoder objective is a reconstruction term that measures how well generated data from $P(X|z)$ (decoder) can represent the input data $X$.

\subsection{Data}

The dataset we use in this research is the one used in the Location-Aware Generative Adversarial Network paper~\cite{oliveira2017learning}. We directly download from the Zenodo website their preprocessed jet images which consist of W boson and QCD jet images~\cite{nachman_benjamin_2017_269622}. We will give a brief summary of the dataset, however, more detail is found in the source paper. The jet images are simulated with Pythia 8.219~\cite{sjstr2007brief,sjostr2006pythia} at 14 TeV and are clustered with FastJet~\cite{cacciari2011fastjet}. The $p_T$ range is set between $250$ GeV $<$ $p_T$ $<$ $300$ GeV and the jets are trimmed and translated so that the subjet with the highest $p_T$ is set at the origin of the image and the second subjet is placed at $-\pi/2$ relative to the $\eta - \phi$ space. The image is graphed with the grid coordinates:

\begin{equation}
\eta \times \phi \in [−1.25, 1.25] \times [−1.25, 1.25]
\end{equation}

This ultimately leads to the production of $25\times25$ images where the pixel intensities represent the total $p_T$. The total $p_T$ corresponds to the equation:
\begin{equation}
p_T = E_{cell} / \cosh(\eta_{cell})
\end{equation}
Where $E_{cell}$ and $\eta_{cell}$ is the energy and pseudorapidity of the calorimeter cells respectively. Once again, a more in-depth detail of the simulation process is found in the paper where we got our dataset. 

\begin{figure}[h!]
  \centering
  \includegraphics[width=0.45\textwidth]{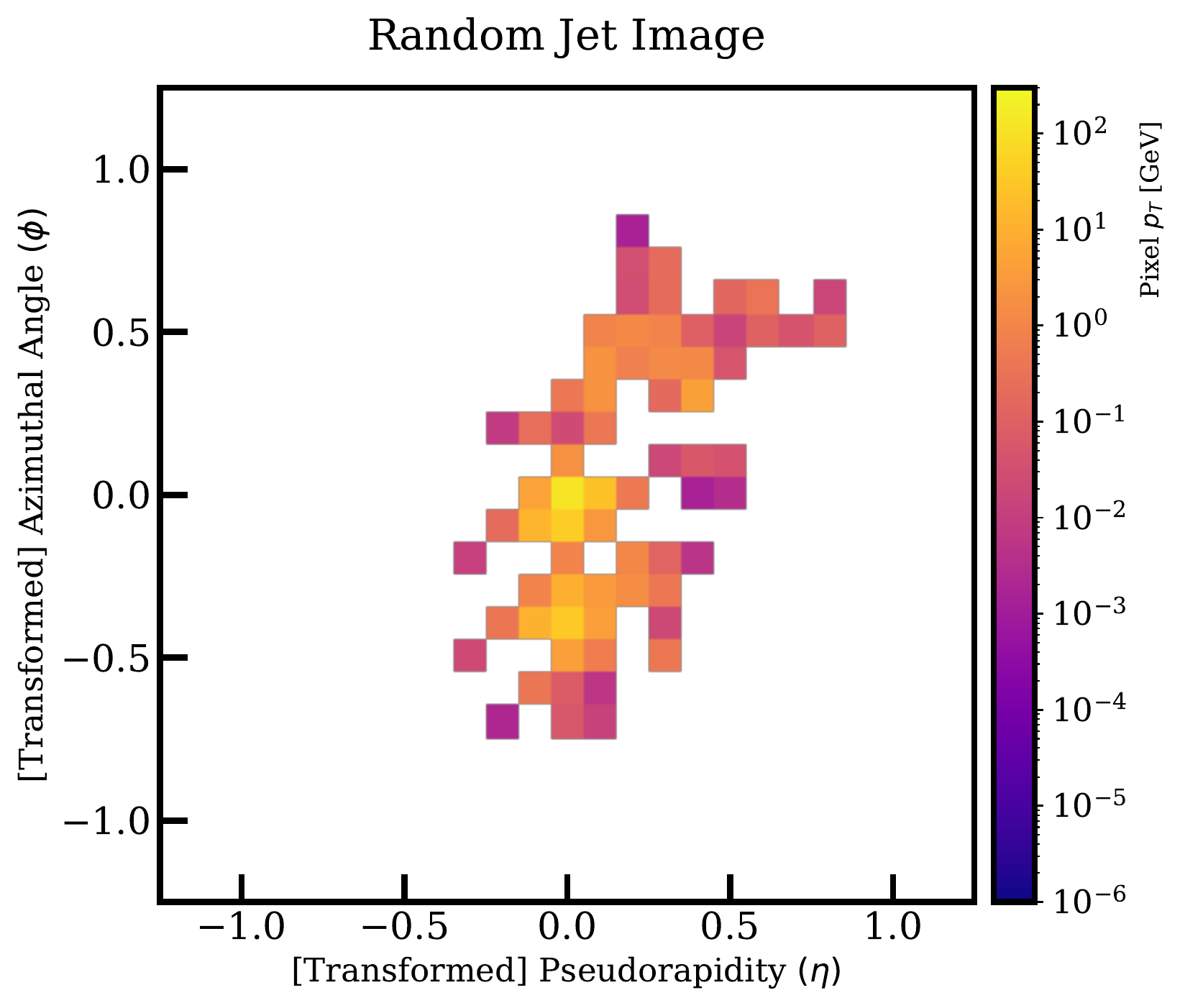}
  \caption{A random jet image.}
\end{figure}

\subsection{Model Architecture}
The model we propose is a vanilla VAE that uses a variety of techniques in VAE literature. The first technique we use borrows from the idea of a Conditional Variational Autoencoder (CVAE) ~\cite{DBLP:conf/nips/SohnLY15}. Since a normal VAE randomly scatters data in the latent space, it is unable to produce images from a certain class. To alleviate this, a CVAE conditions a categorical label to both the input data and the latent space. Because the model we propose uses convolutional layers and it is difficult to combine a three dimensional $25\times25\times1$ image to a one dimensional class label, we use a RepeatVector layer that creates a $25\times25\times2$ image of the categorical class label. This ultimately results in a $25\times25\times3$ input when combined with the input image. This process was important in producing W boson and QCD jet images with distinct features.

With the $25\times25\times3$ image, we pass through an encoder that consists of three convolutional layers with filters of 32, 64, 128 respectively. A kernel size of 3 is used for all layers, the padding is same for all layers, and strides of 2 are used only for the second and third convolutional layers of the encoder. We flatten the last convolutional layer of the encoder and pass the flattened image into the latent space. The categorical label is conditioned to the latent space, which is then passed through a dense layer of 6272 and reshaped to a dimension of $7\times7\times128$. This then goes through a decoder that consists of two deconvolutional layers with filters of 64, 32 respectively, kernel sizes of 3, and strides of 2. For the last convolutional layer, we put a convolutional layer with 1 filter, valid padding, and a kernel size of 4 to return to a $25\times25\times1$ image. All convolutional layers use LeakyReLU activation functions except for the last layer, which uses relu as recommended in~\cite{oliveira2017learning}. 

To further amplify the quality of the jet images, we take the sum of the KL divergence, feature perceptual and Bernoulli loss as our loss function. Normally, a VAE uses the KL divergence loss paired with some simple distance metric such as the mean squared error. We use the feature perceptual loss as our distance metric loss instead due to the higher quality images it produces~\cite{hou2016deep, johnson2016perceptual}. 

The feature perceptual loss takes the squared difference between the hidden features of the input and output image of the VAE. The hidden features are calculated by a pretrained convolutional neural network that is excluded from the optimization process of the model. The loss function is mathematically described as:

\begin{equation}
\begin{gathered} 
{\Lagr_{perceptual}^i} = \frac{1}{2C^iW^iH^i}\sum_{c=1}^{C^i}\sum_{w=1}^{W^i}\sum_{h=1}^{H^i}(\Phi(\hat{X})^i - \Phi(X)^i) ^2 \\
{\Lagr_{perceptual}} = \sum_i(w_i \times \Lagr_{perceptual}^i)
\end{gathered}
\end{equation}

Where $\Phi(X)$ and $\Phi(\hat{X})$ are the predicted features, taken from the $ith$ layer in the convolutional neural network, of the input and output images and $C, W, H$ is the channel, width, and height of the images respectively. The total loss is calculated by combining the feature perceptual loss for $i$ layers in the convolutional neural network with a weight, denoted by $w_i$, for each loss. The weight can be adjusted to put emphasis on certain layers when calculating the loss. For simplicity, we set the weight for each perceptual loss as 1. 

We found that the feature perceptual loss was a crucial component in capturing the sparse radiation patterns and the prongs of the jet images. Because of this, the use of locally-connected layers, which was recommended in~\cite{oliveira2017learning}, was not needed. In addition to the feature perceptual loss, we add a Bernoulli loss function. Because the Bernoulli loss requires gray scale images, we normalize the images by dividing them by 100. We also transform both the input and output image to 625 dense layers when calculating this loss. We found that using this loss helped produce slightly better sparsity. 

This makes the total loss function of the VAE:

\begin{equation}
{\Lagr_{VAE}} = \Lagr_{perceptual} + \Lagr_{KL} + \Lagr_{Bernoulli}
\end{equation}

The convolutional neural network for calculating the feature perceptual loss uses three convolutional layers with three dense layers. We tested how various kernel sizes and filters affected the performance of the VAE to see which classifier architecture was optimal. We found that using relatively large kernel sizes paired with 32 filters in the first convolutional layer was able to teach the VAE how to produce the central region of the jets. We ultimately settled on using a kernel size of 11 with 32 filters. The second layer, as our investigations showed, helped in producing the sparse radiation patterns of the jet images. We found that using small kernel sizes of 3 or 5 and 64 filters was able to produce sparse jet images. We ultimately decided to settle on a kernel size of 5. We found that the third layer, which had 64 filters and a kernel size of 5, was not necessary in calculating the feature perceptual loss and even worsened performance. 

Each convolutional layer uses strides of two and are followed by a Dropout layer. The LeakyReLU activation is also used for each convolutional layer. The convolutional layers in the classifier are followed by three dense layers of 128, 64, and 1, where the last layer has a sigmoid activation function. The model is trained for 50 epochs with a batch size of 100 and uses the Adam optimizer. The loss for the classifier is the binary crossentropy loss. 

\begin{equation}
BCE = (y)(\minus\log(y_{pred}) + (1-y)(\minus\log(1-y_{pred}))
\end{equation}

To clarify, we use the same dataset to train the classifier as the one used to train the VAE. This means the classifier is doing a binary classification on W boson and QCD jet images.

We trained multiple VAE models and found that the latent size was also crucial in producing sparse jet images. We found that a large latent size of 10-15 produced jet images with good sparsity and prominent jet prongs. Anything below 6 didn't produce as much sparsity but the jet prongs were preserved. Anything higher than 15 produced jet images with unrealistic sparsity levels. This ultimately makes sense because the latent size is where information is stored, therefore, a bigger latent size will store more information while a smaller one will limit information. In this case, more information is equivalent to capturing more sparsity in the jet images. We found that a latent size of 12 was a good dimension for  $25\times25\times1$ images. The latent size should ultimately vary depending on the dimension of the input data and the type of data being used. 

We build all models, including the feature perceptual classifier, in Keras~\cite{chollet2015keras} with a Tensorflow backend~\cite{tensorflow2015-whitepaper}. We use the Adam optimizer and train the VAE for 12 epochs with a batch size of 100. The model is trained in Kaggle which uses a NVidia K80 GPU.

\begin{figure}[h!]
  \centering
  \includegraphics[width=1\textwidth]{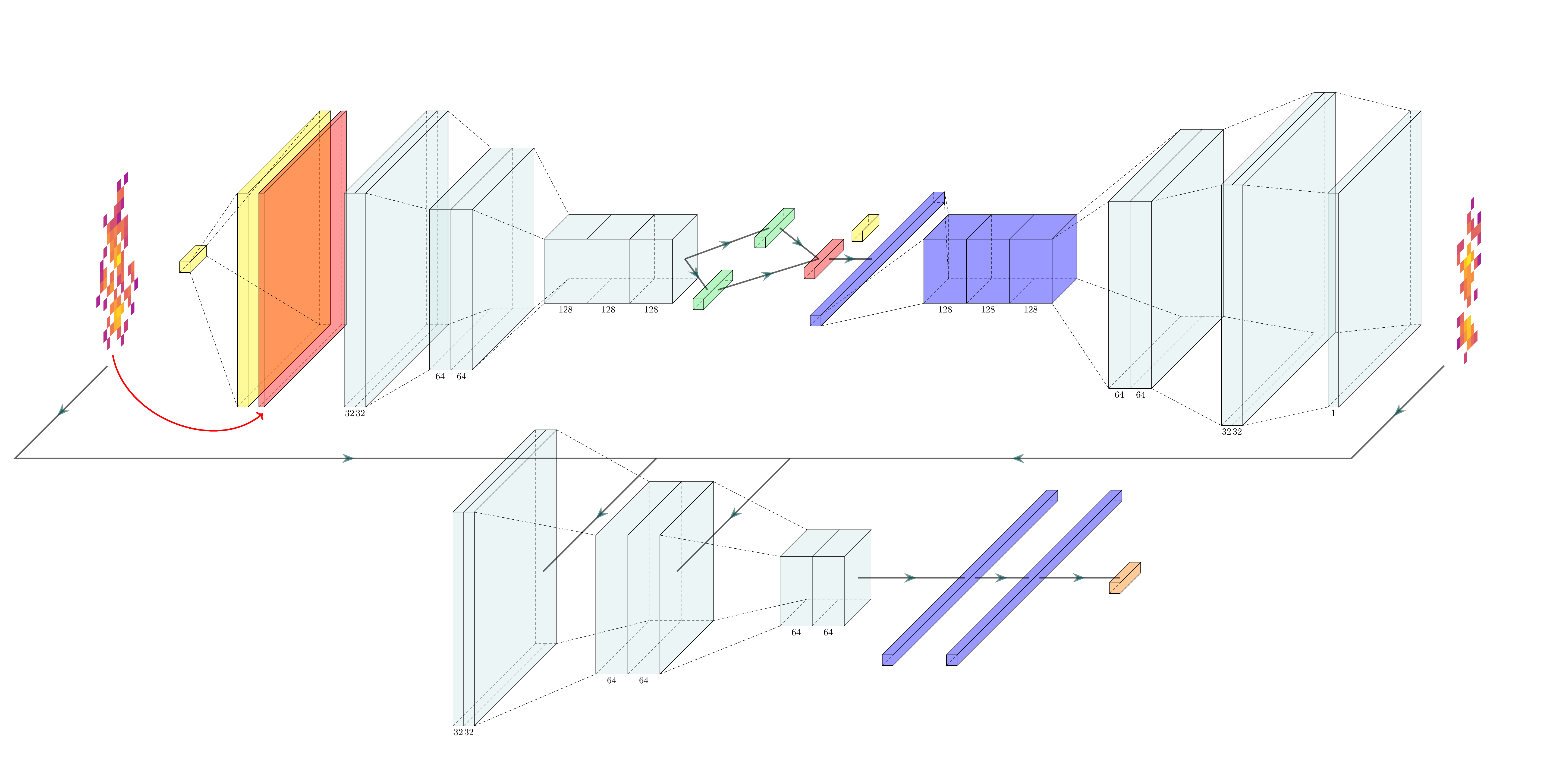}
  \caption{The architecture of our VAE model. The VAE starts off by passing the categorical label through a RepeatVector layer. This 3D categorical image label is then combined with the input image, resulting in a $25\times25\times3$ input. This input is passed through the encoder, which then goes through a sampling layer. The latent space is concatenated to the label, passed through a dense $6272$ layer, reshaped to $7\times7\times128$, then passed through the decoder. The input and output images are compared with the hidden features of the first two layers in the classifier.}
  \label{VAEA}
\end{figure}

\section{Model Assessment}

\begin{figure}[h!]
  \centering
  \hspace*{-1cm}\includegraphics[width=0.46\textwidth]{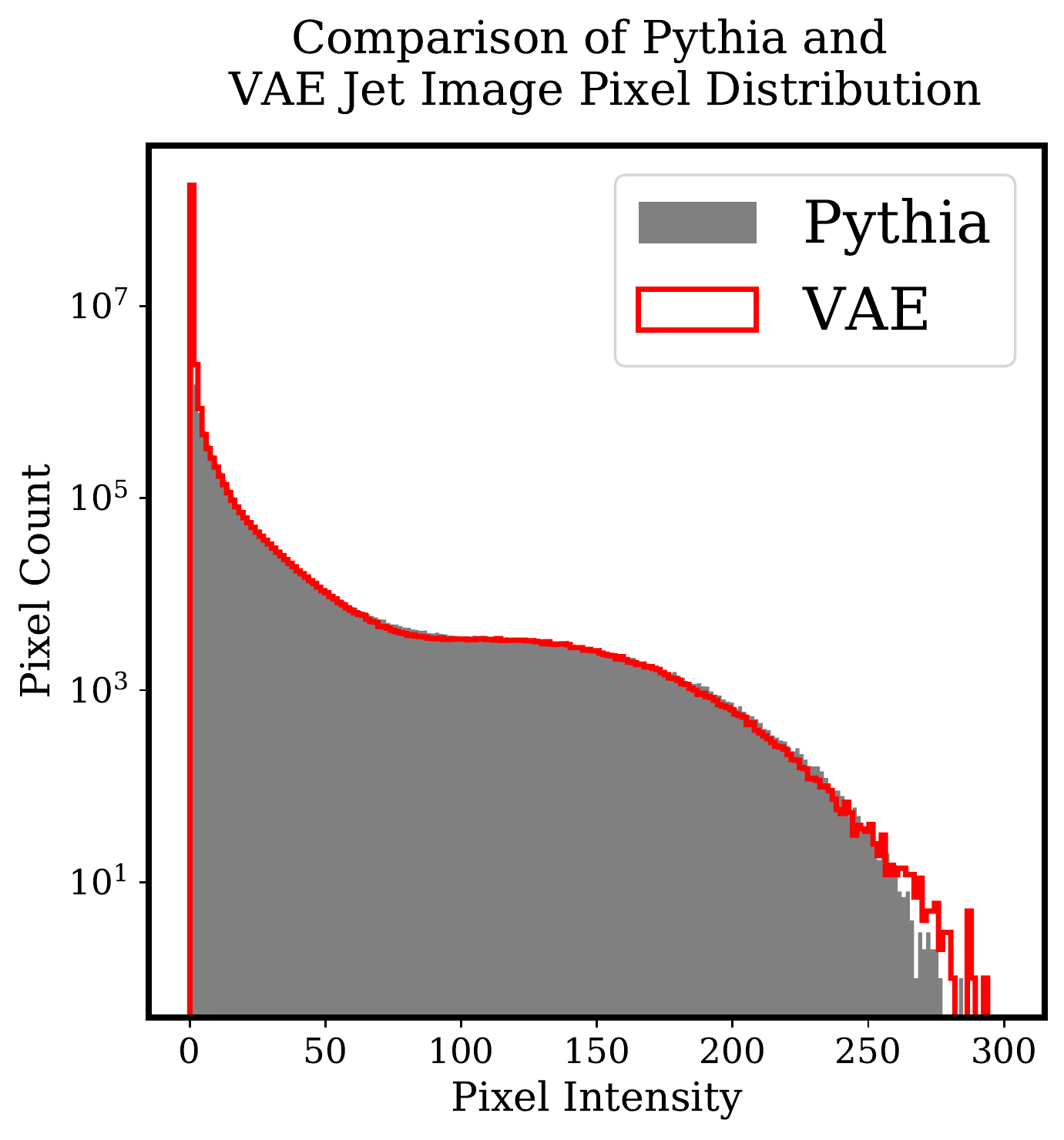}
  \caption{Pixel distribution of Pythia and VAE jet images.}
  \label{Pixel Intensity}
\end{figure}

For our assessment, we compare 300,000 VAE and Pythia jet images. We begin our VAE assessment by graphing a histogram of the pixel intensities of both Pythia and VAE jet images in Figure \ref{Pixel Intensity}. The pixel distribution helps assess whether or not the VAE can successfully learn a wide variety of pixel intensities. In this case, the VAE is shown to explore high and low levels of pixel intensities and is able to match the overall pixel distribution of the Pythia jet images. Similarly, we can measure the difference between Pythia and VAE by taking the mean squared error of their jet images. We also calculate the MSE between Pythia and the Location-Aware Generative Adversarial Network (LAGAN)~\cite{de_oliveira_luke_2017_400706} jet images. Both values are shown in Table \ref{MSE}. The VAE is shown to have a slightly smaller difference with Pythia jet images in comparison to the LAGAN.

\begin{table}[h!]
\large
\centering
\caption{Mean Squared Error (MSE) of Pythia jets with VAE and LAGAN jets.}
\begin{tabular}[t]{p{5.5cm} cc}
\hline
Model & MSE\\ [0.2cm]
\rowcolor{red!20}
\textbf{VAE} & 17.27\\ 
\textbf{LAGAN} & 19.29\\
\hline
\end{tabular}
\label{MSE}
\end{table}%

\begin{figure}[h!]
\begin{subfigure}{.32\textwidth}
  \centering
  \includegraphics[width=1\linewidth]{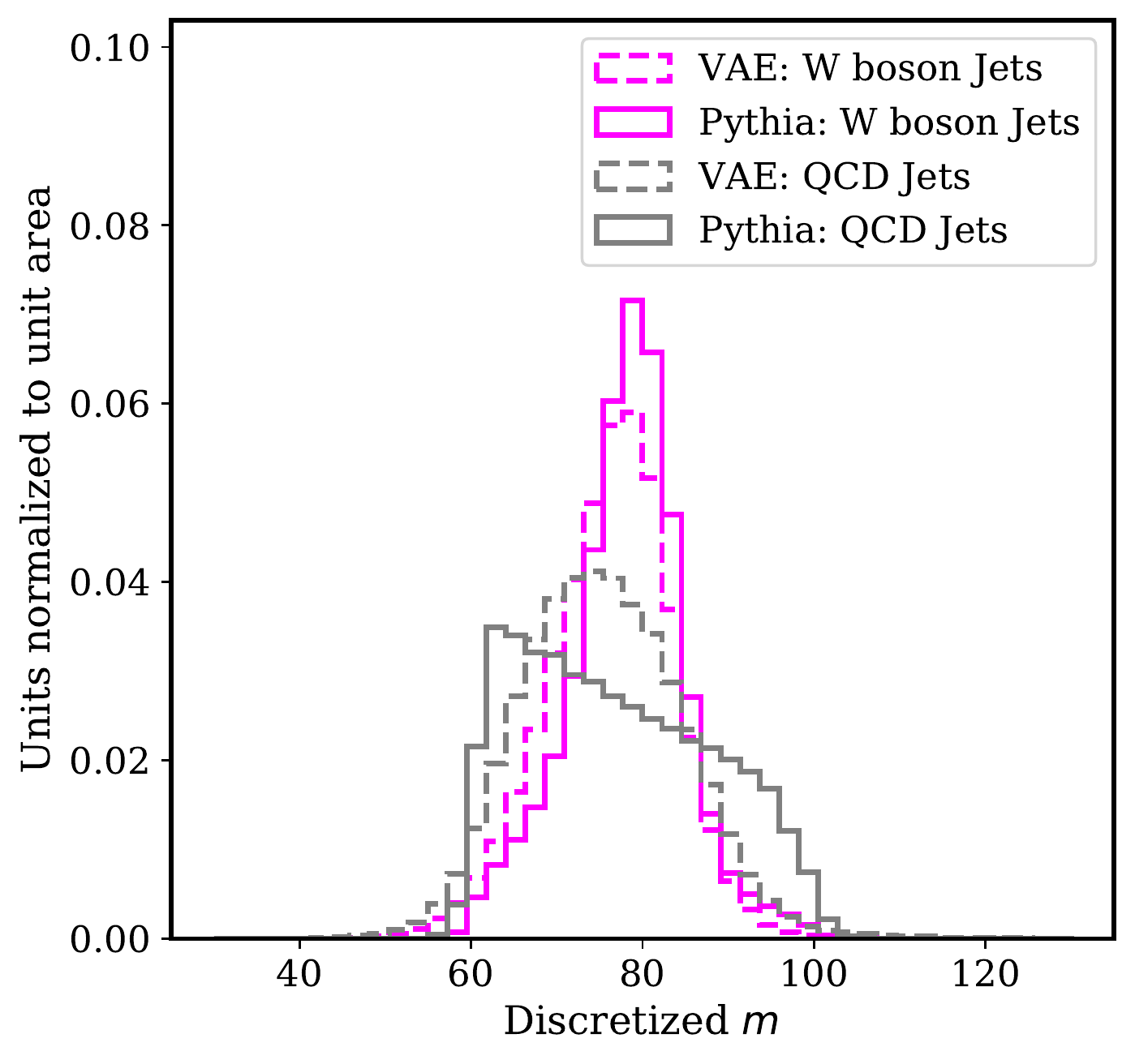}  
\end{subfigure}
\begin{subfigure}{.32\textwidth}
  \centering
  \includegraphics[width=1\linewidth]{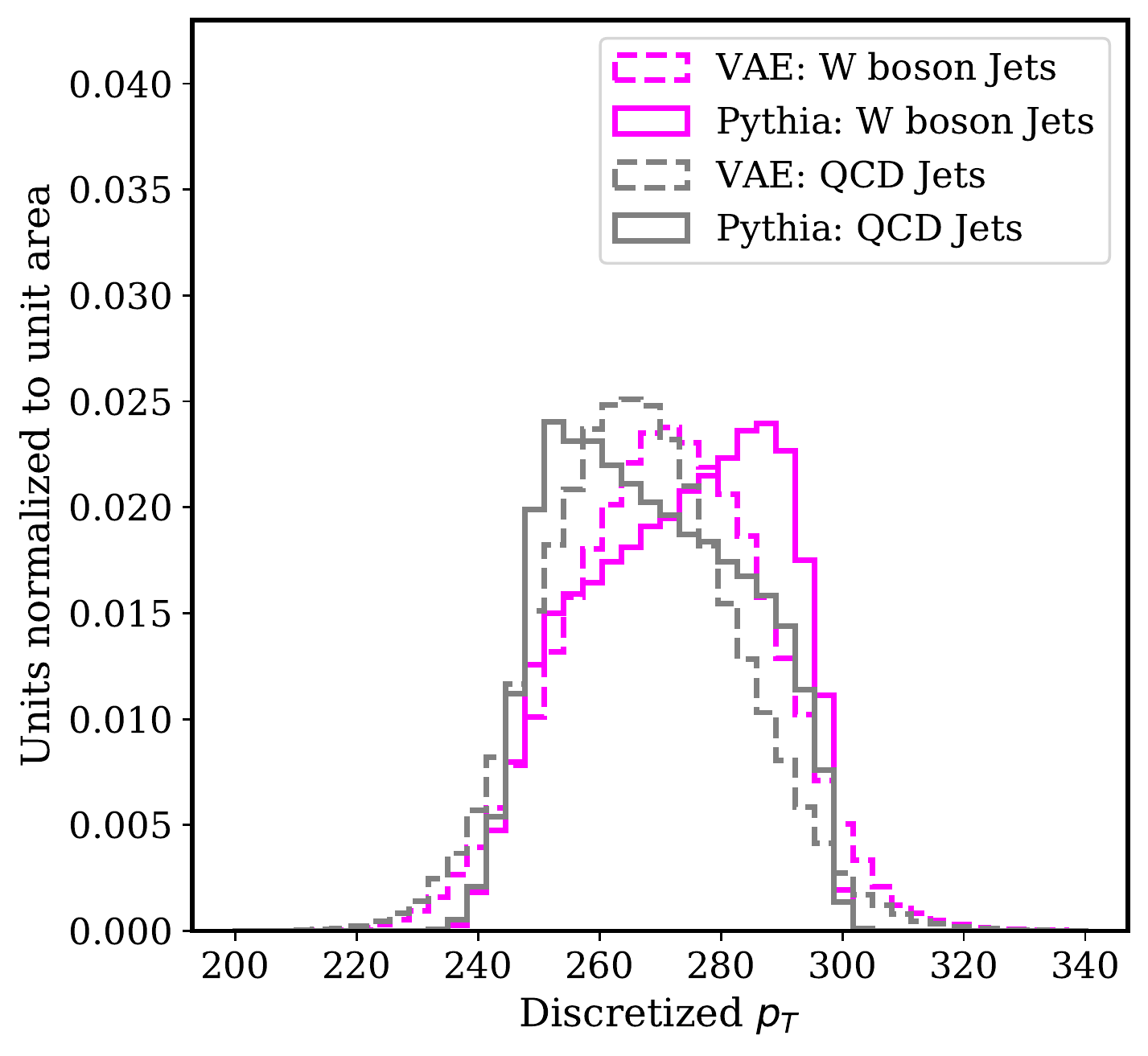}
\end{subfigure}
\begin{subfigure}{.32\textwidth}
  \centering
  \includegraphics[width=1\linewidth]{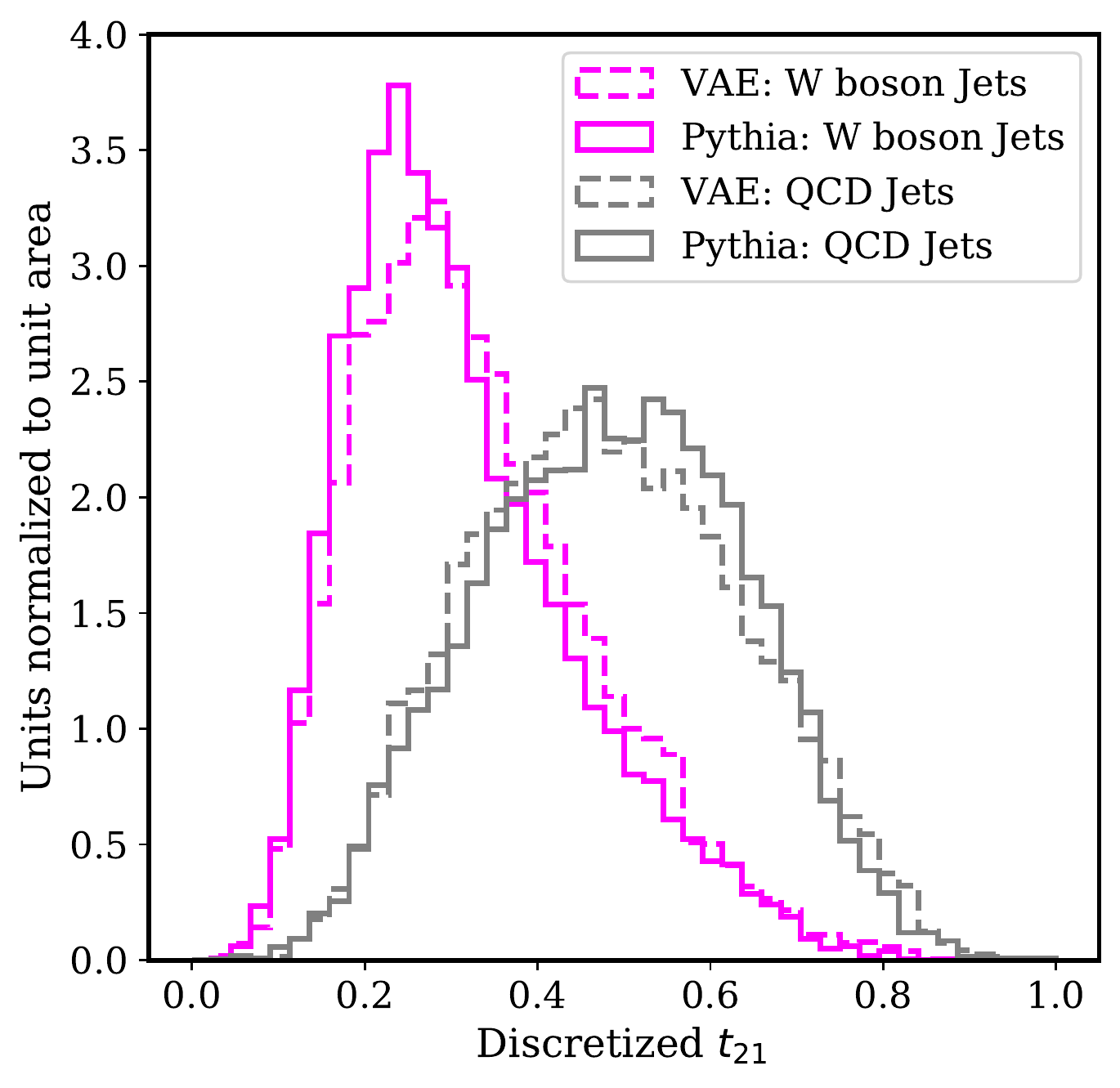}  
\end{subfigure}
\caption{Mass, $p_t$, and N-subjettiness distributions for Pythia and VAE jet images.}
\label{Distributions}
\end{figure}

\begin{figure}[h!]
  \centering
  \includegraphics[width=1\textwidth]{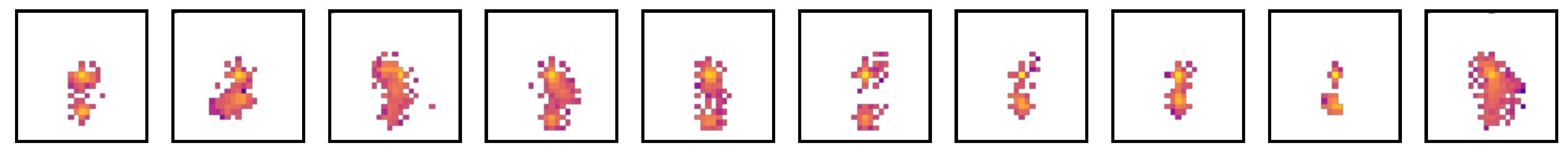}
  \caption{10 random VAE jet images. The images highlight that the VAE, despite using the feature perceptual loss, can still produce slightly blurry images.}
  \label{Random VAE Images}
\end{figure}

Another useful way to assess our VAE model is to see whether or not its jet images can recreate some of the jet observables of the Pythia jet images. In this study, we calculate the mass, $p_T$, and N-subjettiness \cite{thaler2010identifying} of both Pythia and VAE jet images and compare them in Figure \ref{Distributions}. The VAE is shown to match the general shape of the Pythia jet observables, however, it struggles in certain areas. This is especially noticeable in the mass of the QCD jet images. This is further highlighted in Table \ref{Mu_Sigma}, where the mean and standard deviation of the VAE distributions are shown. The VAE does a slightly poorer job in matching the Pythia distribution than the LAGAN. Despite the effectiveness of the feature perceptual loss, the VAE still produces blurrier images, especially for sparse features. This can be seen in Figure \ref{Random VAE Images} where the jet images are shown to be ``smudged". Additional modifications, such as adding additional loss functions and constraints, may be able to mitigate this problem. 

\begin{table}[ht]
\centering
\begin{tabular}{@{\extracolsep{4pt}} llc|cc|cccc}
\toprule   
{} & \multicolumn{2}{c}{Mass} & \multicolumn{2}{c}{$p_T$}  & \multicolumn{2}{c}{$t_{21}$}\\
 \cmidrule{2-3} 
 \cmidrule{4-5} 
 \cmidrule{6-7} 
 Model & $\mu$ & $\sigma$ & $\mu$ & $\sigma$ & $\mu$ & $\sigma$\\ 
\midrule
\rowcolor{magenta!70!white!30}
\textbf{Pythia (W boson)} & 78.19 & 6.77 & 273.12 & 14.91 & 0.31 & 0.13 \\ 
\vspace{0.03cm}
\textbf{VAE (W boson)} & 76.16 & 7.34 & 271.08 & 14.51 & 0.33 & 0.14 \\ 
\vspace{0.3cm}
\textbf{LAGAN (W boson)} & 79.64 & 7.24 & 276.27 & 14.81 & 0.28 & 0.09 \\
\rowcolor{gray!20}
\vspace{0.03cm}
\textbf{Pythia (QCD)} & 76.91 & 11.01 & 268.64 & 14.82 & 0.49 & 0.15 \\ 
\vspace{0.03cm}
\textbf{VAE (QCD)} & 75.29 & 9.40 & 266.03 & 14.08 & 0.48 & 0.16 \\ 
\vspace{0.03cm}
\textbf{LAGAN (QCD)} & 77.62 & 13.30 & 267.06 & 19.44 & 0.51 & 0.13 \\ 
\bottomrule
\end{tabular}
\caption{Mean and standard deviation of jet observable distributions.} 
\label{Mu_Sigma}
\end{table}

\begin{figure}[h!]
\begin{subfigure}{.5\textwidth}
  \hspace*{3em}
  \centering
  \includegraphics[width=0.8\linewidth]{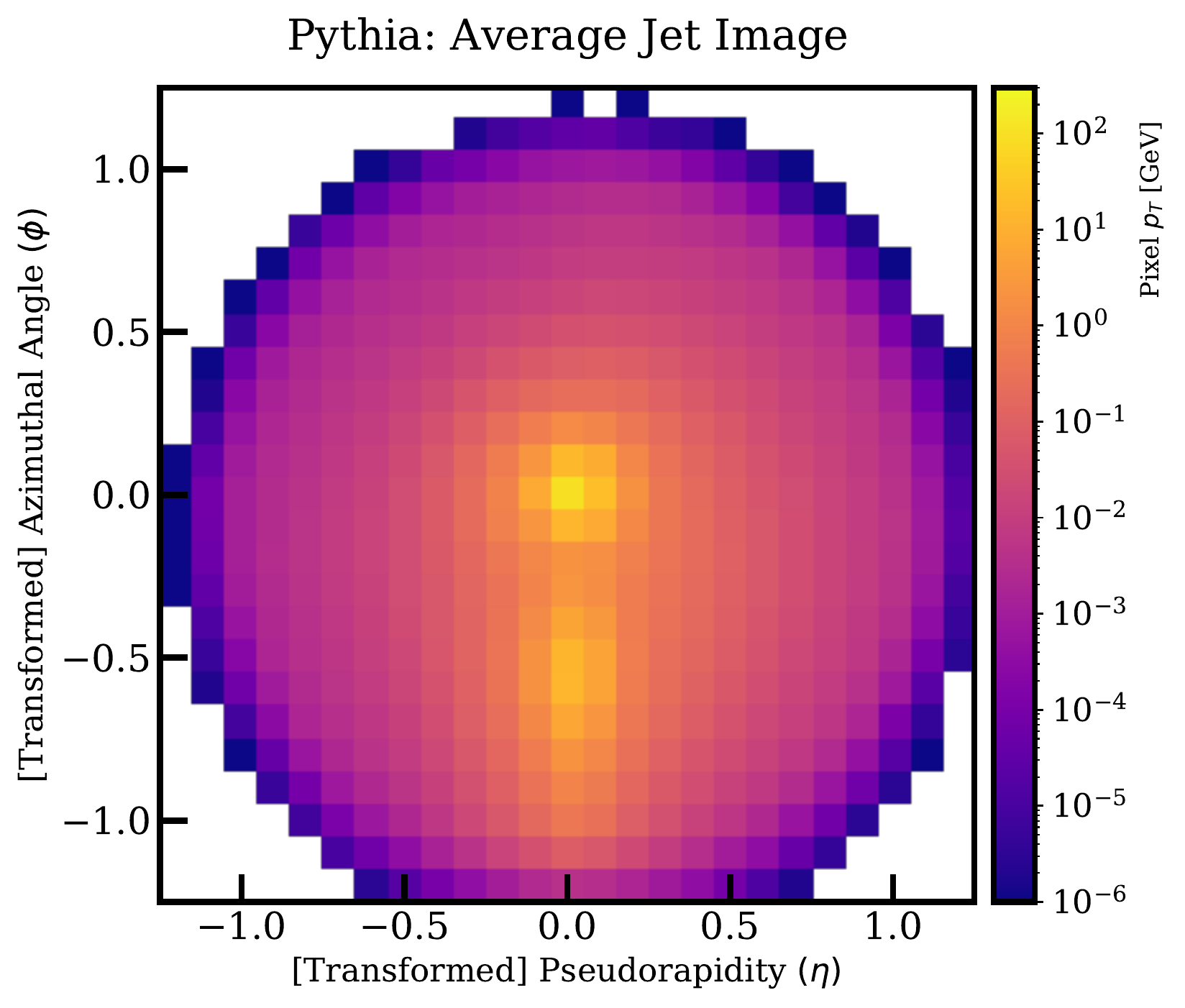}
\end{subfigure}
\begin{subfigure}{.5\textwidth}
  \hspace*{-3em}
  \centering
  \includegraphics[width=0.8\linewidth]{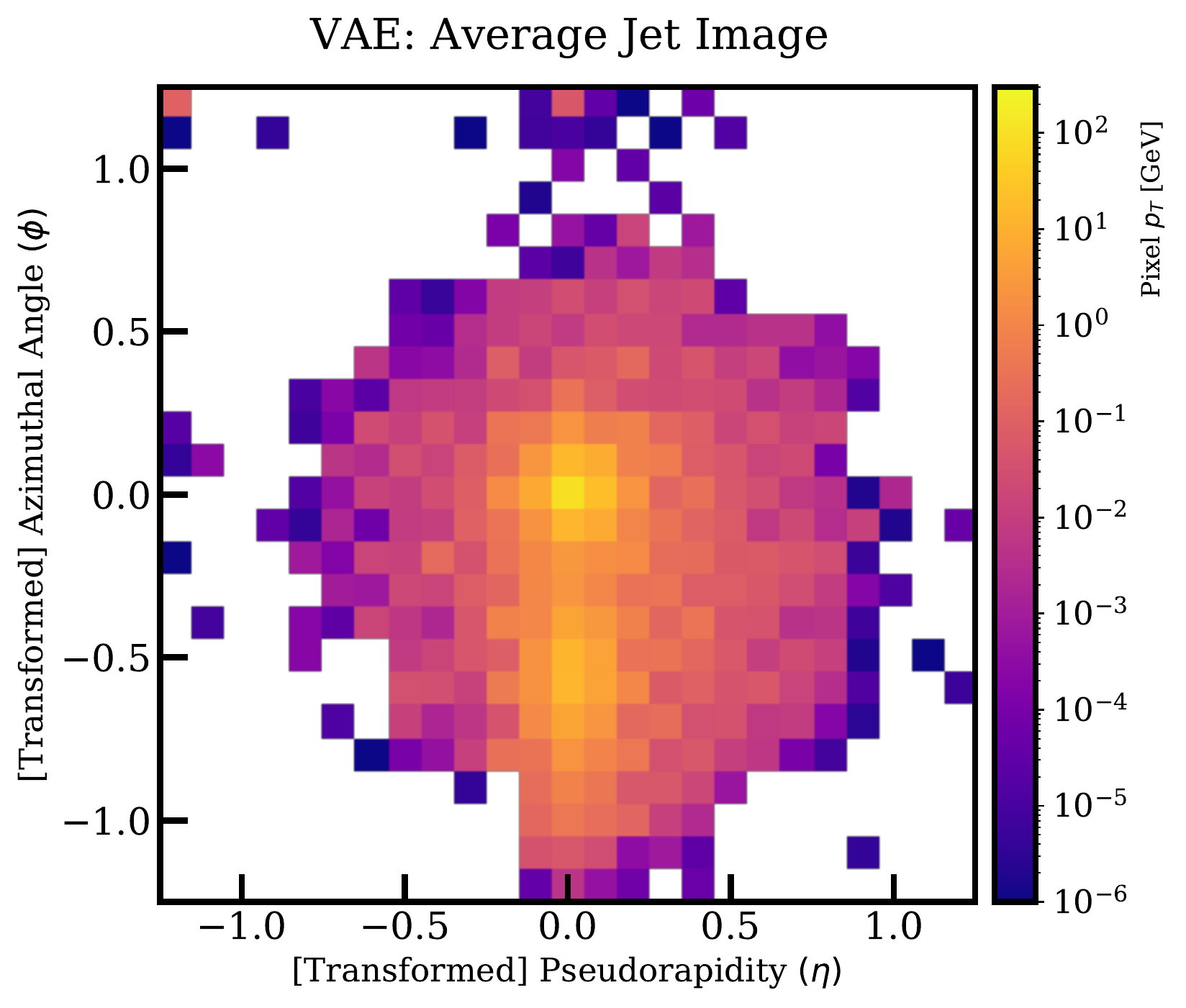}  
\end{subfigure}
\caption{Average jet image for Pythia and VAE jet images.}
\label{Average Images}
\end{figure}

\begin{figure}[h!]
\begin{subfigure}{.5\textwidth}
  \hspace*{3em}
  \centering
  \includegraphics[width=0.75\linewidth]{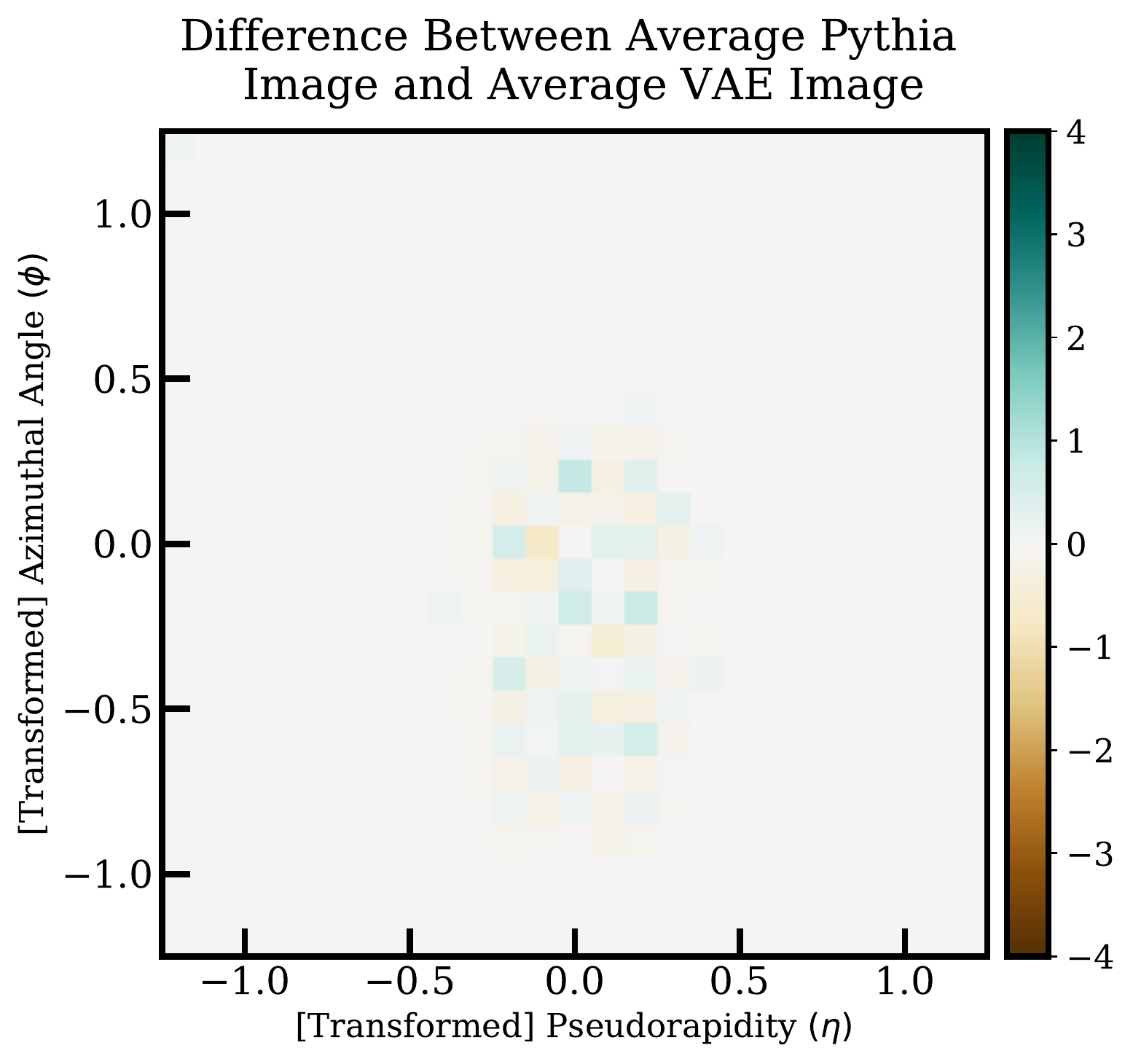}
\end{subfigure}
\begin{subfigure}{.5\textwidth}
  \hspace*{-3em}
  \centering
  \includegraphics[width=0.75\linewidth]{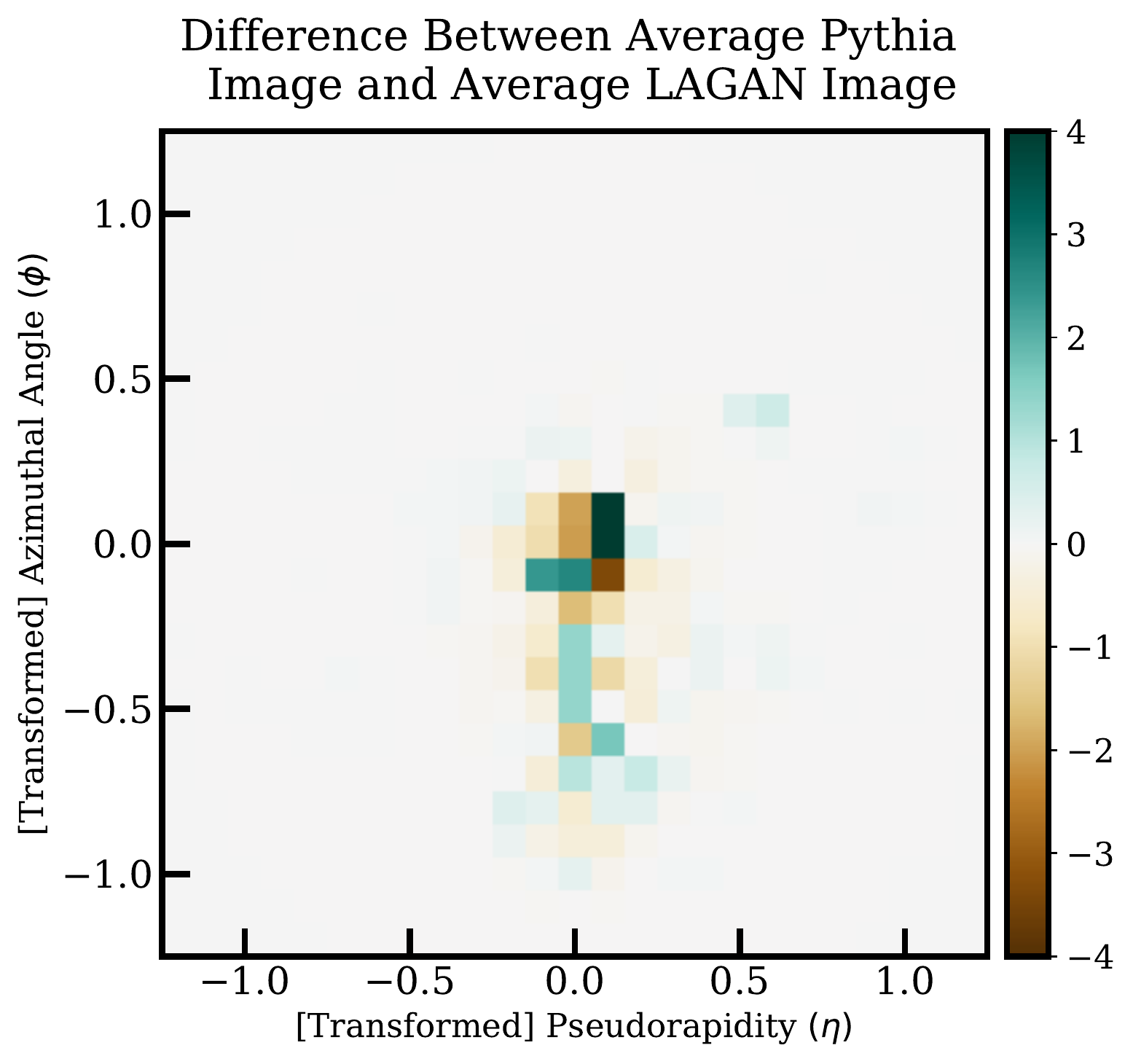}
\end{subfigure}
\caption{Difference between the average Pythia jet image with the average VAE and LAGAN jet image.}
\label{PCC Average Images}
\end{figure}

To provide a more visual assessment of what kind of jet images the VAE produces, we take the average jet image for both Pythia and VAE jet images in Figure \ref{Average Images}. The average VAE jet image is shown to reproduce the jet constituents well along with some of the outer radiation. One problem outlined by the average VAE jet image is that some of the radiation towards the bottom left is missing. This may happen due to the nature of the feature perceptual loss. When the classifier is calculating the loss between the input and output image, parts of the hidden features may be empty due to the use of same padding in the first and second layers of the classifier. To avoid this, it may be best to avoid same padding in the classifier altogether. Despite this problem, we support the claim that the VAE is able to reproduce the central region of the Pythia jet images by taking the difference between the average Pythia and average VAE jet image in Figure \ref{PCC Average Images}. We also include the difference between the average Pythia jet image with the average jet image from the LAGAN. 

The central component of the VAE jet image is shown to be similar to the central component of the Pythia jet image due to the little variation there is between the two images. This is strongly correlated to the feature perceptual loss and the latent size we used. Without the use of the feature perceptual loss, we found that the VAE did a significantly poorer job in producing the central component of its jet images. Using big kernel sizes and a relatively small number of filters for the first layer of the classifier used for the feature perceptual loss also helped reinforce the learning of the central region. In addition to the feature perceptual loss, the latent size we chose was also crucial for producing the central component of the jets. Having too big of a latent size negatively affected the intensities towards the center. We did see, however, that the central region was well produced even with latent space sizes smaller than 10. This is most likely the case because the central region of the jet images is a more prominent feature, therefore, the VAE will learn this feature much more consistently. To compare our VAE with previous generative models, we also include the difference between the average Pythia jet image and the average LAGAN jet image. The LAGAN is shown to struggle more in producing the central component of its jets due to the strong positive and negative correlations shown in the image. To quantitatively measure Figure \ref{PCC Average Images}, we calculate the Structural Similarity Index (SSIM) of the average Pythia image with the average VAE and LAGAN image in Table \ref{SSIM}. The table ultimately demonstrates that the VAE can reproduce the central region of the jet images significantly better than the LAGAN.

\begin{table}[h!]
\large
\centering
\caption{{Structural Similarity Index (SSIM) of the average Pythia jet image with the average VAE and LAGAN jet image. The closer the SSIM is to 1, the better the performance.}}
\begin{tabular}[t]{p{5.5cm} cc}
\hline
Model & SSIM\\ [0.2cm]
\rowcolor{red!20}
\textbf{VAE} & 0.94\\ 
\textbf{LAGAN} &0.85\\
\hline
\end{tabular}
\label{SSIM}
\end{table}%

\begin{figure}[h!]
\hspace*{0.4em}
\begin{subfigure}{.32\textwidth}
  \centering
  \includegraphics[width=1\linewidth]{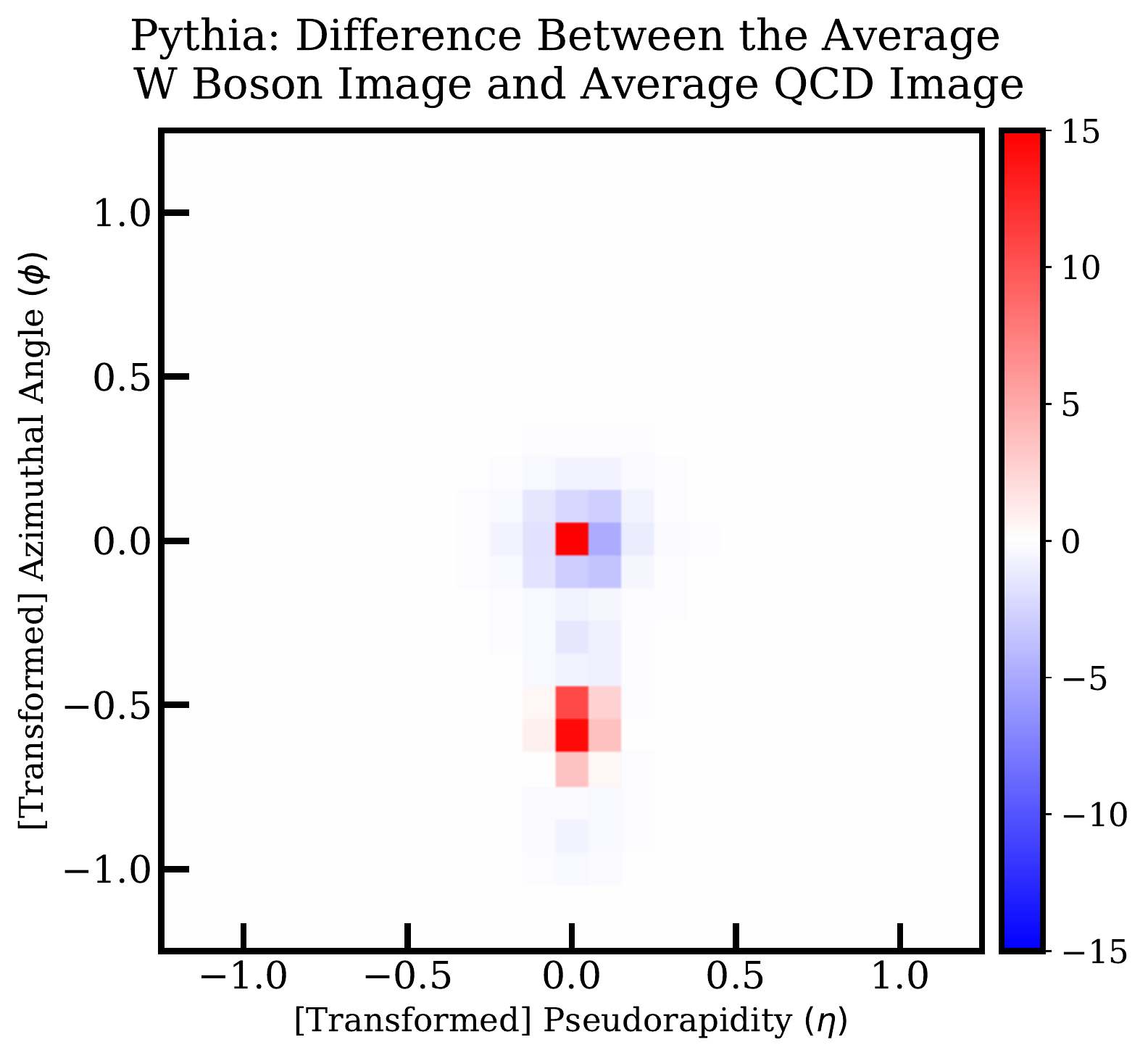} 
\end{subfigure}
\begin{subfigure}{.32\textwidth}
  \centering
  \includegraphics[width=1\linewidth]{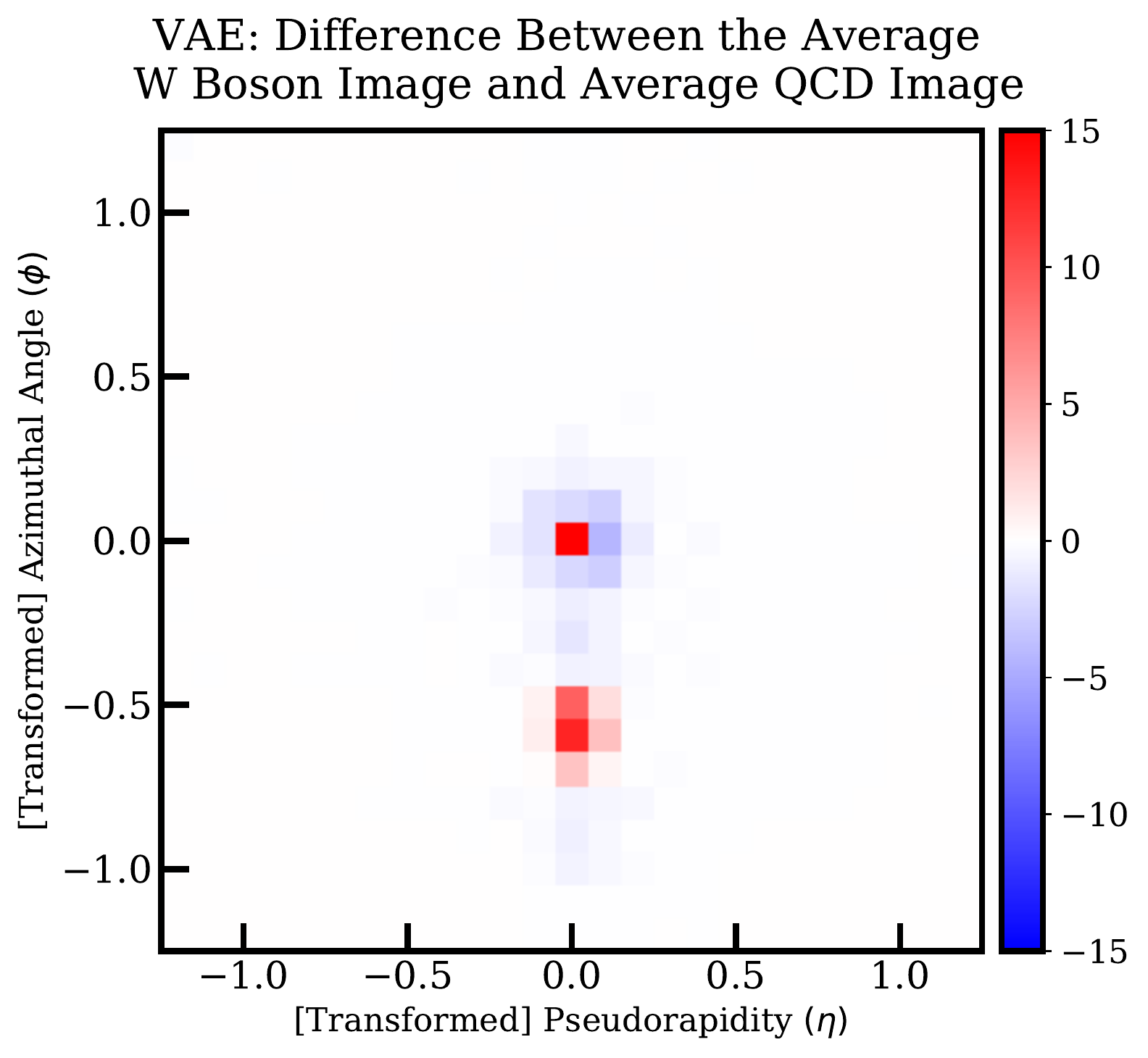}  
\end{subfigure}
\begin{subfigure}{.32\textwidth}
  \centering
  \includegraphics[width=1\linewidth]{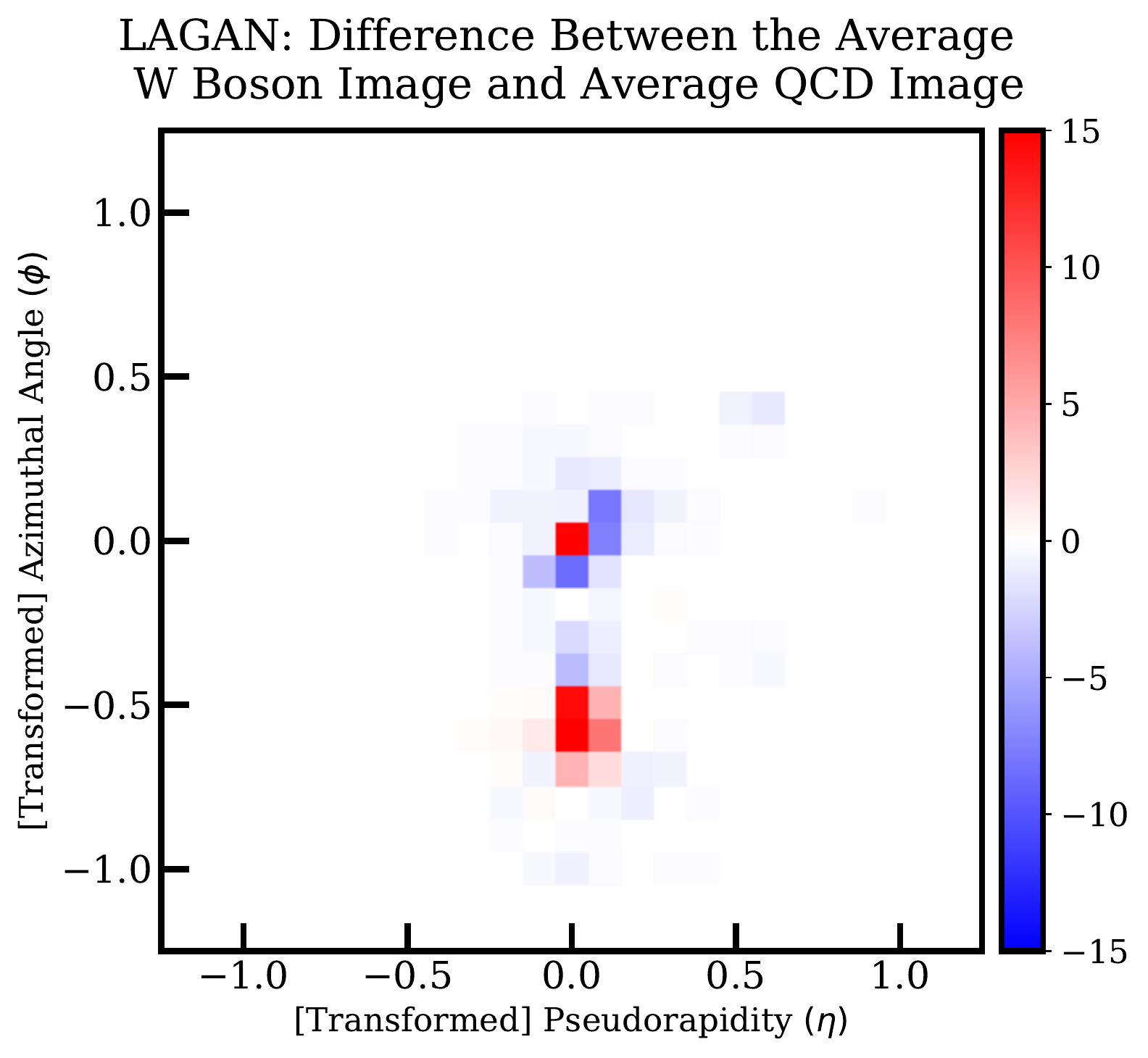}
\end{subfigure}
\caption{Difference between the average W boson and QCD jet image for Pythia, VAE, and LAGAN.}
\label{PCC_W_QCD}
\end{figure}

To evaluate the VAE's ability to generate different classes of images, we take the difference between the average W boson and QCD jet image produced by the VAE. This is an important evaluation because the features between the two classes of images should be noticeable. In this case, we want to see the outer radiation of the QCD jet images and the jet prongs from the W boson jet images. This was one of the shortcomings of the LAGAN because the model overestimated the importance of certain features in its jet images. To compare the distinguishability of W boson and QCD jet images, we take the difference between the average W boson and average QCD jet images for Pythia, VAE, and LAGAN in Figure \ref{PCC_W_QCD}.

The VAE is shown to have almost identical differences in its W boson and QCD jet image with the W boson and QCD jet images of Pythia. The radiation pattern of the QCD jet image and the jet prongs of the W boson jet image is visible in the VAE. This contrasts the LAGAN, where the radiation of the QCD jet image is more prominent than it should be. 

\begin{figure}[h!]
  \centering
  \hspace*{-0.5cm}\includegraphics[width=0.46\textwidth]{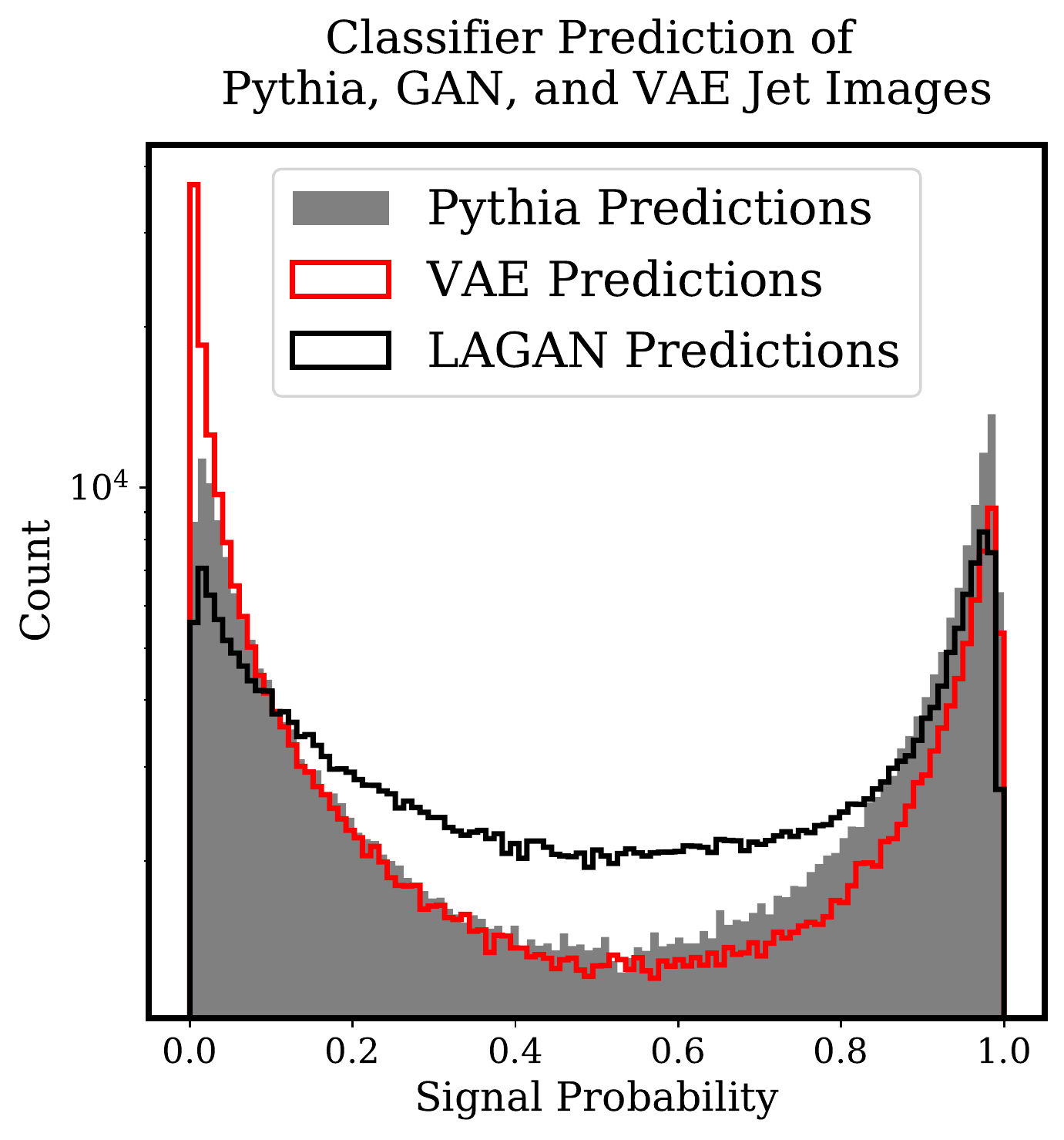}
  \caption{Classifier class probability predictions for Pythia, VAE, and LAGAN jet images. 0 represents a background prediction and 1 represents a signal prediction. The VAE is shown to produce QCD jet images that are easier to classify. The overall distribution, however, is shown to match a good majority of the Pythia distribution.}
  \label{Classifier}
\end{figure}

We further explore the similarity of W boson and QCD jet images by taking the predicted class probabilities, calculated by the feature perceptual loss classifier, of the jet images in Figure \ref{Classifier}. This classifier was trained with Pythia jets only, so the ultimate goal is to match the Pythia curve. 0 represents a prediction of background and 1 represents a prediction of signal. The classifier predicts Pythia, VAE, and LAGAN jets.

We can see that that the distribution of the Pythia jet images has a nice symmetrical distribution. The VAE is shown to have slightly higher predictions for background and slightly lower predictions for signal. Although the VAE curve is not perfect, it is ultimately able to match a good majority of the Pythia predictions. The LAGAN predictions are symmetrical, however, there is more uncertainty when classifying its jet images. This is demonstrated with the higher counts towards the center of the distribution. 

\section{Exploring the Latent Space}

\begin{figure}[h!]
  \centering
  \includegraphics[width=0.8\textwidth]{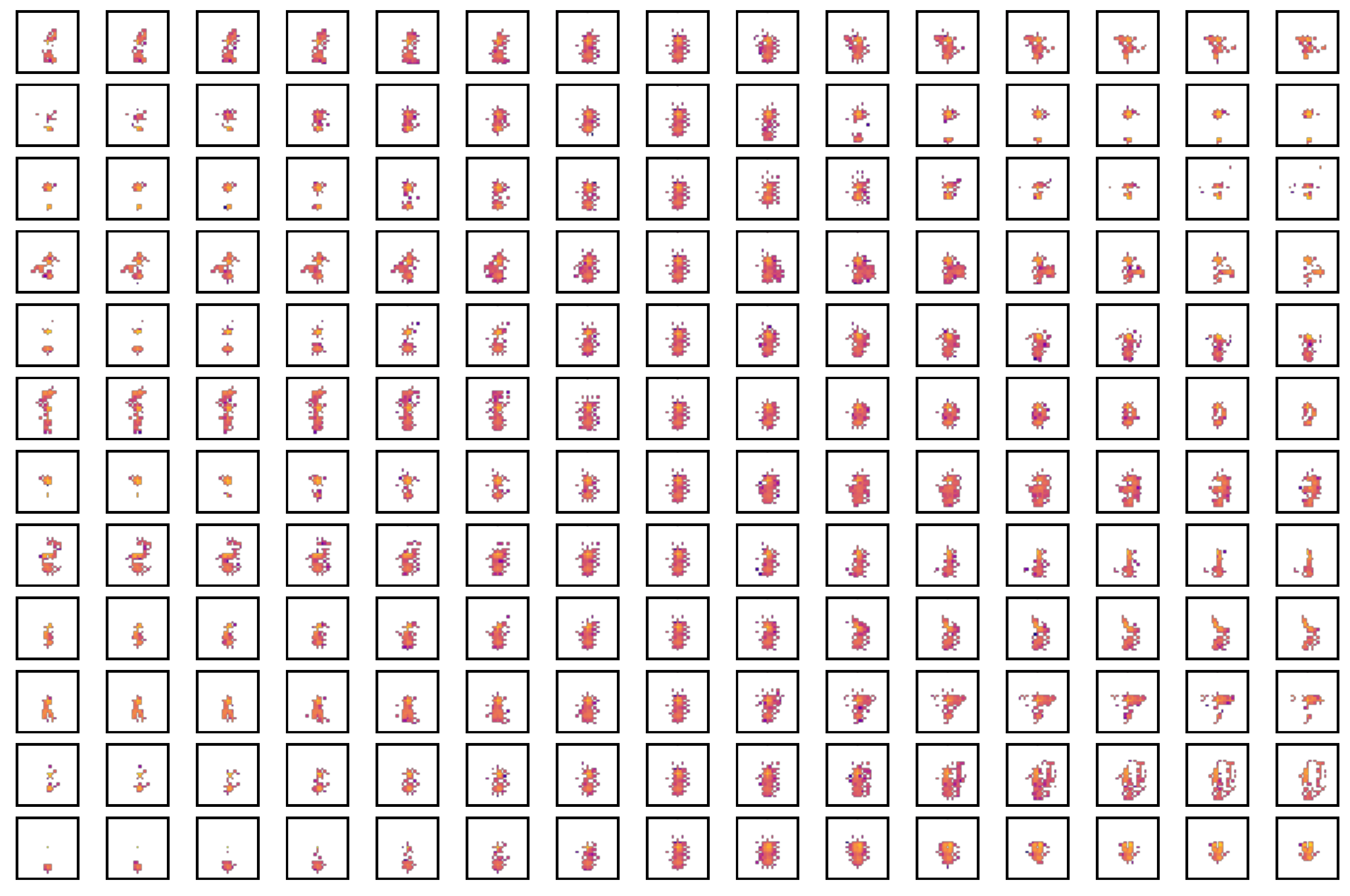}
  \caption{Linear Interpolation of QCD jet images.}
  \label{LIQCD}
\end{figure}

\begin{figure}[h!]
  \centering
  \includegraphics[width=0.8\textwidth]{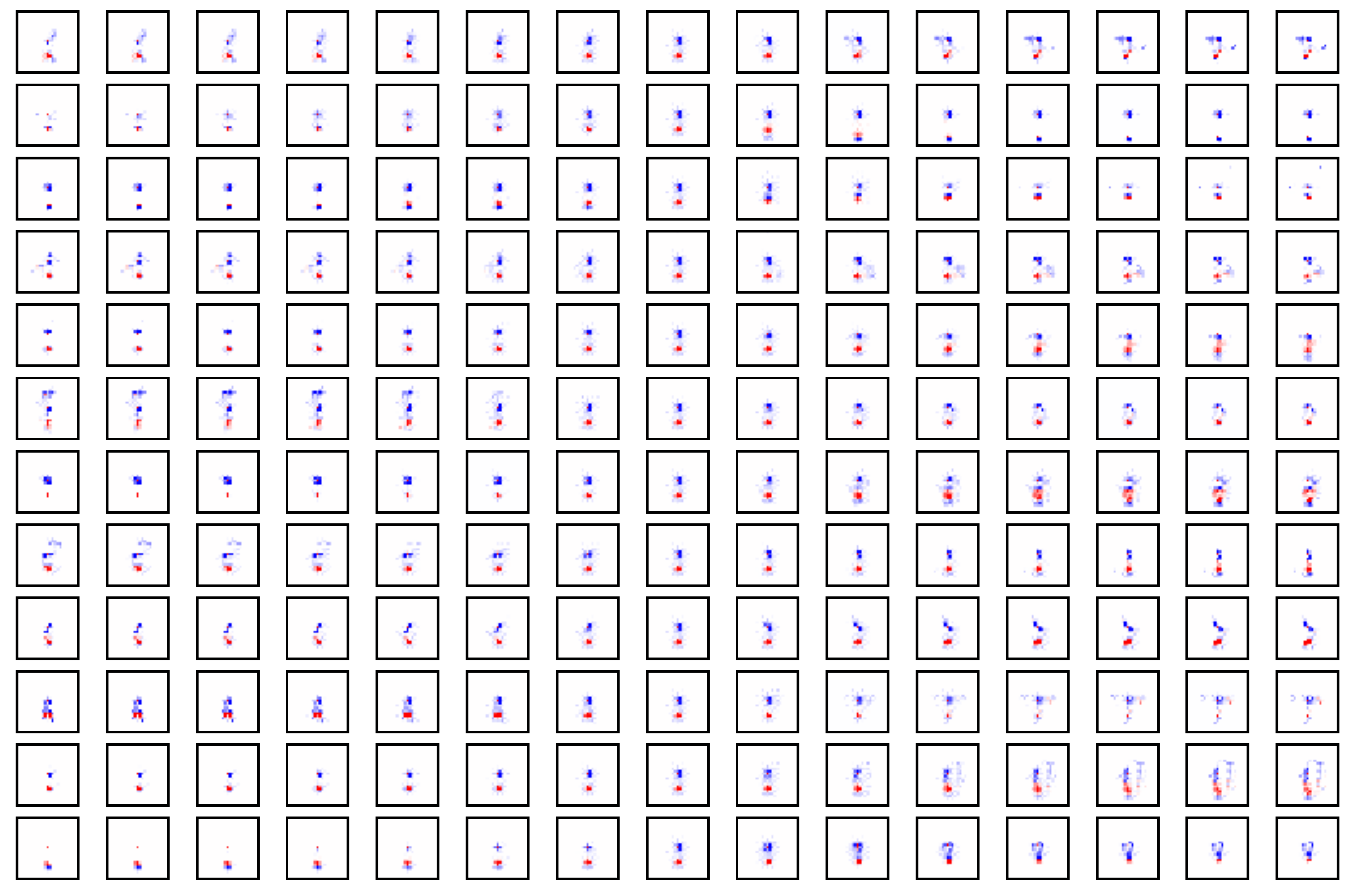}
  \caption{Difference in Linear Interpolation between W boson and QCD jet images.}
  \label{LIPCC}
\end{figure}

\begin{figure}[h!]
  \centering
  \includegraphics[width=0.8\textwidth]{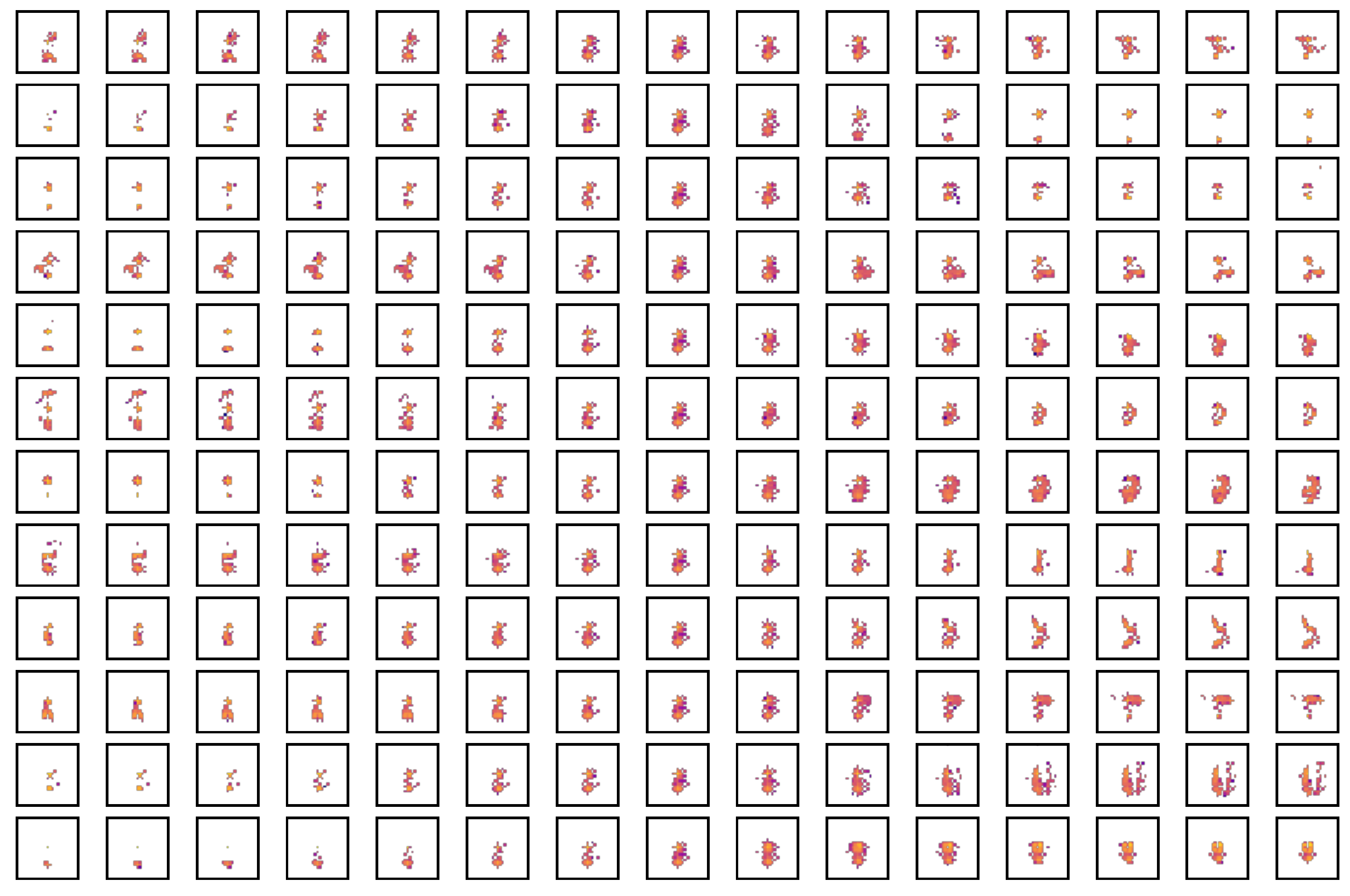}
  \caption{Linear Interpolation of W jet images.}
  \label{LIW}
\end{figure}

A key feature of the variational autoencoder is its latent space. This is where are all the information of the VAE is stored, therefore, it is important to see how the latent space affects what kind of jet images are produced. We start this section by displaying what are known as linear interpolations. A linear interpolation is a series of images generated by a VAE that travels linearly through the Gaussian distribution used to simulate data. In this case, the distribution has a mean of 0 and a standard deviation of 1. The linear interpolation ultimately helps visualize what kind of features the VAE learns. We begin by showing a linear interpolation of QCD and W boson jets in Figures \ref{LIQCD} and \ref{LIW} and the difference between the two classes of images in Figure \ref{LIPCC}. Each column represents a different latent space value (1-12) in descending order. 

The VAE is shown to have learned a diverse range of features in the latent space. Two examples include the slants found in the 10th latent space value and the growing radiation patterns in the 7th latent space value. These linear interpolations help us not only visualize more specific features that the VAE is learning, but it also demonstrates what kind of images are produced for each latent space value. A more detailed study is found later in this section.

\begin{figure}[h!]
  \centering
  \includegraphics[width=0.88\textwidth]{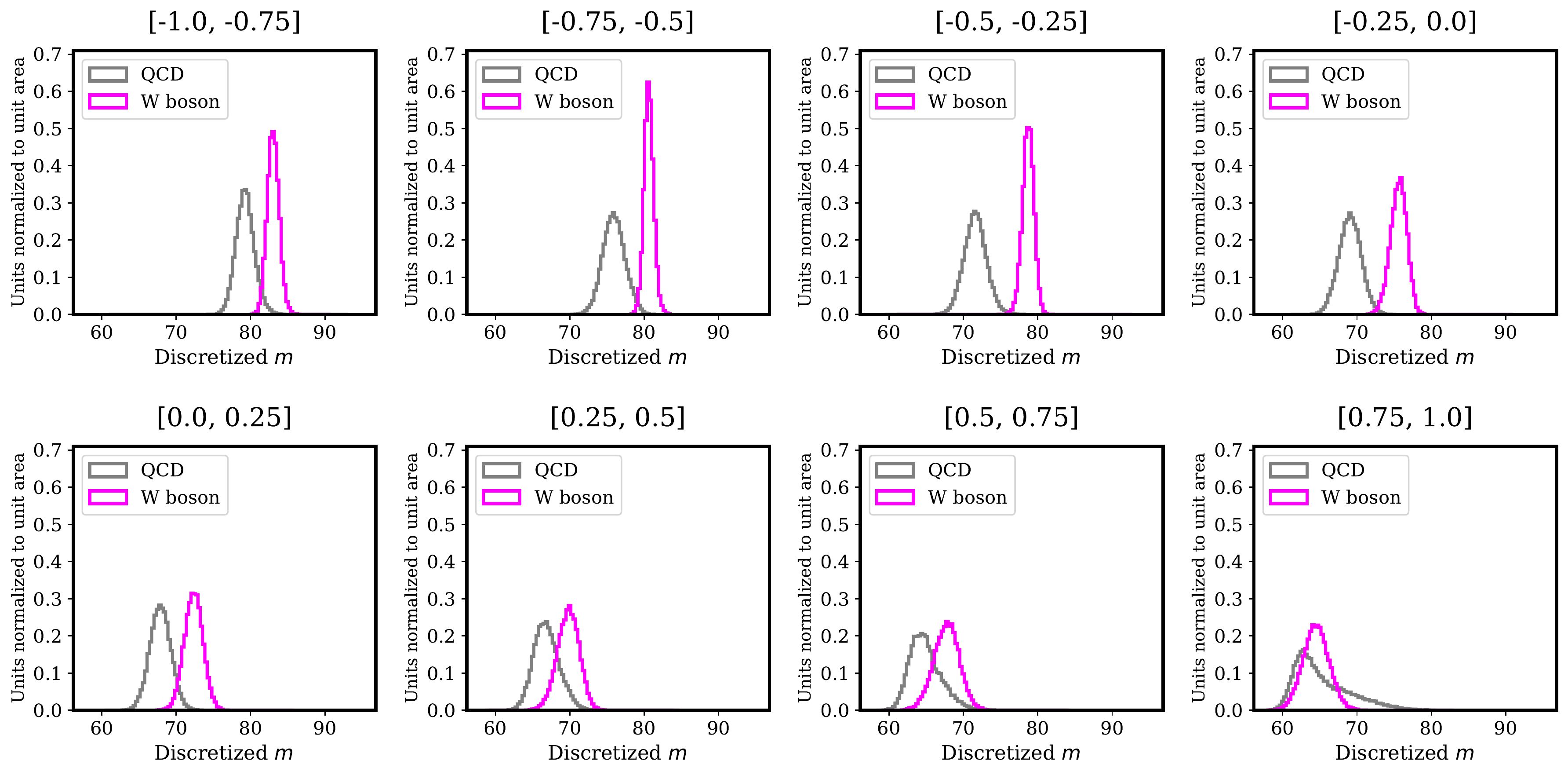}
  \caption{Mass distributions for various intervals in the $N(0,1)$ distribution.}
  \label{LIMD}
\end{figure}

\begin{figure}[h!]
  \centering
  \includegraphics[width=0.88\textwidth]{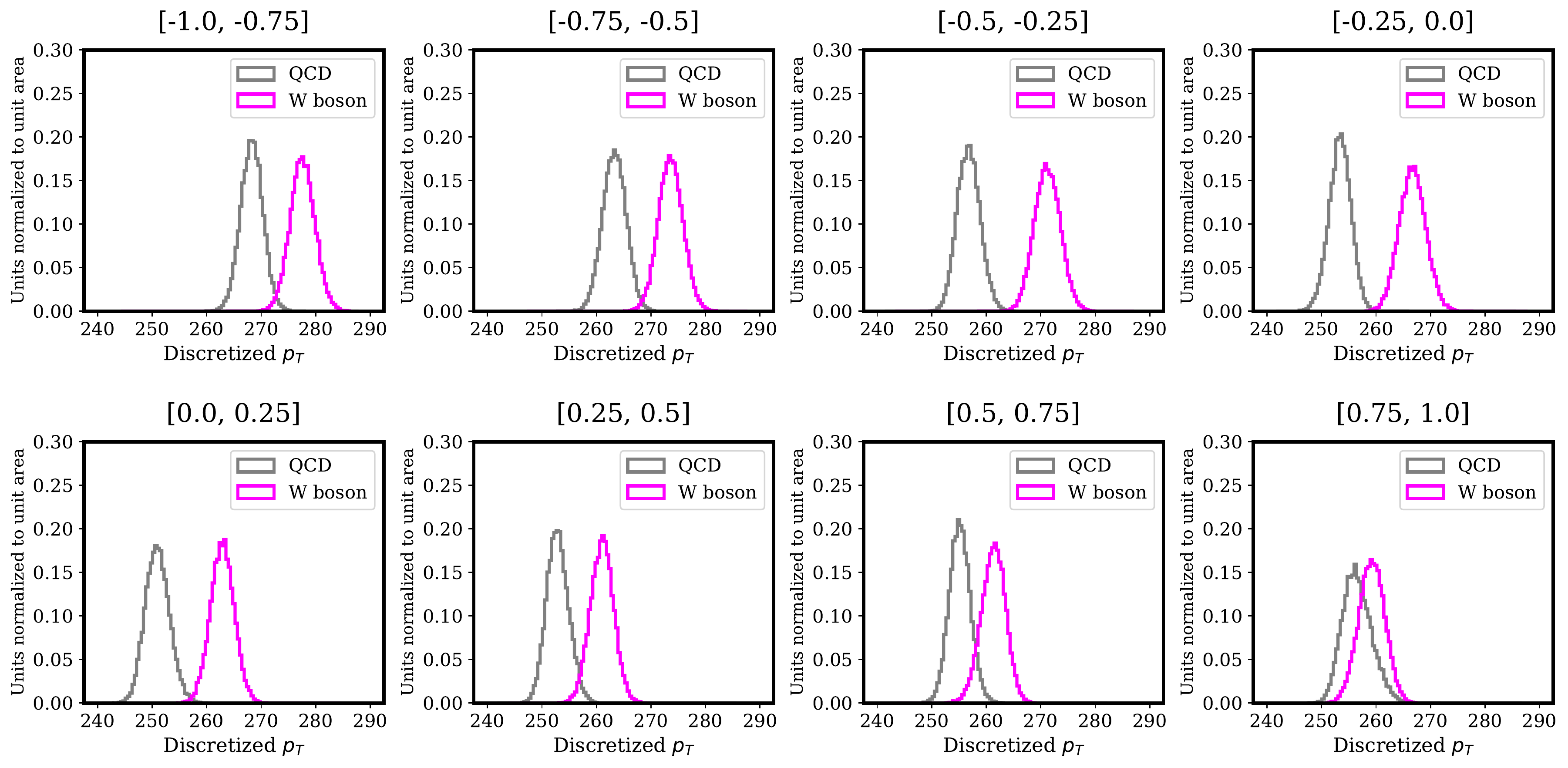}
  \caption{$p_T$ distributions for various intervals in the $N(0,1)$ distribution.}
  \label{LIPD}
\end{figure}

\begin{figure}[h!]
  \centering
  \includegraphics[width=0.88\textwidth]{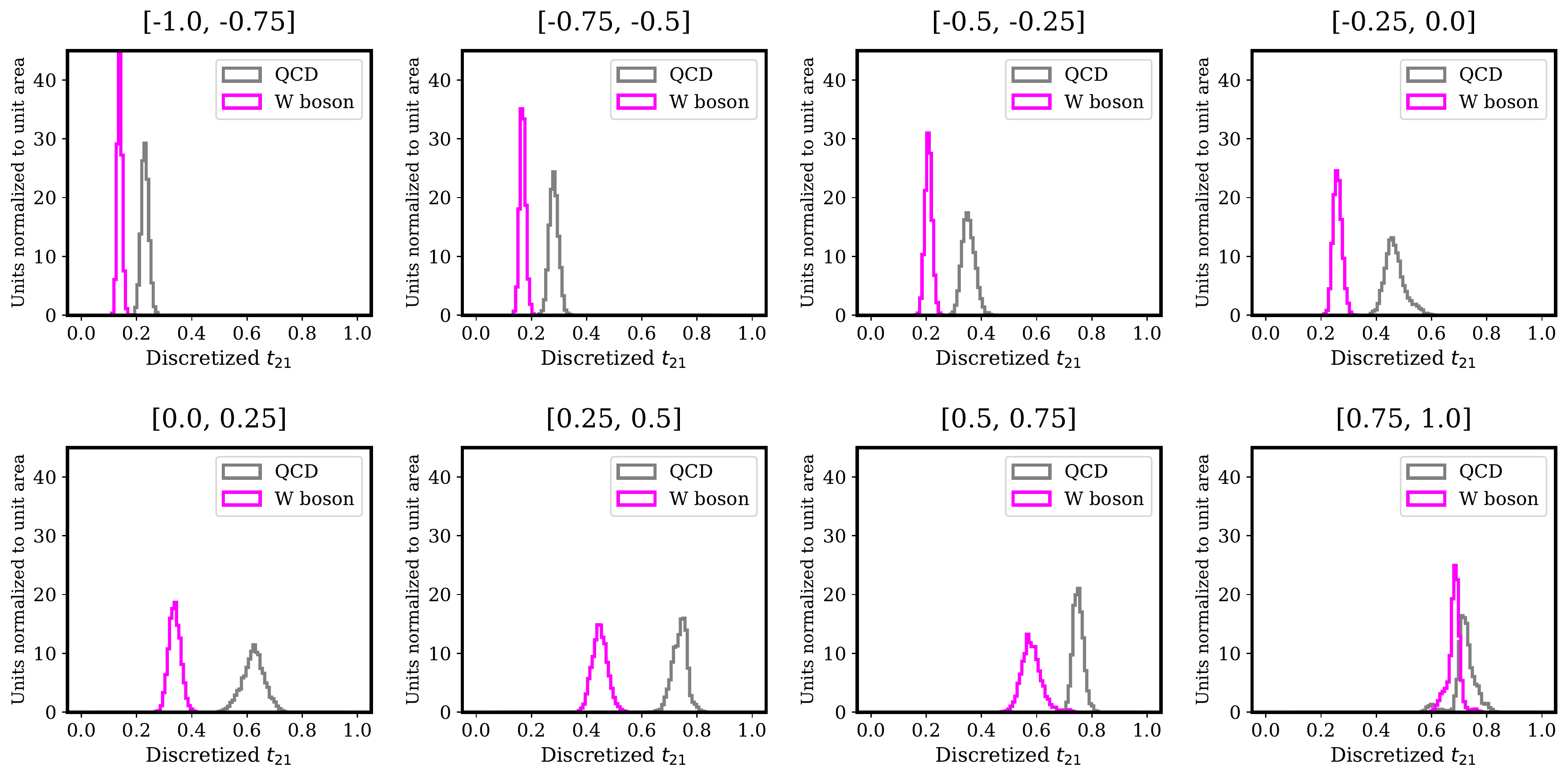}
  \caption{N-subjettiness distributions for various intervals in the $N(0,1)$ distribution.}
  \label{LIPT}
\end{figure}

\begin{figure}[h!]
  \centering
  \includegraphics[width=1\textwidth]{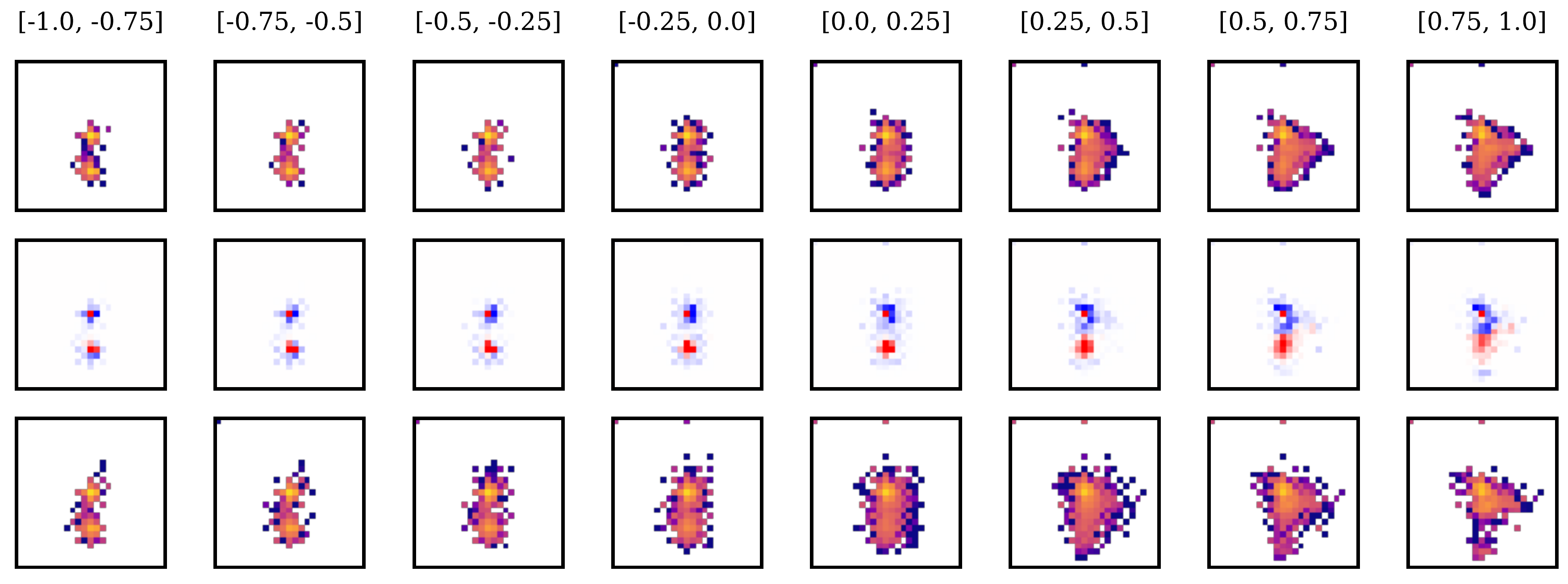}
  \caption{Average jet image for different intervals in the N(0,1) distribution. The top column represents the average W boson jet image, the bottom represents the average QCD jet image, and the center column is the difference between the two images. We can see that the outer radiation of the QCD (blue) and the prongs of the W boson (red) jet image become more prominent as they approach 1.}
  \label{LIAJI}
\end{figure}

We provide a more detailed analysis on how the features of the jets change by seeing how the mass, $p_T$, and N-subjettiness curves in Figures \ref{LIMD}, \ref{LIPD}, and \ref{LIPT} change in different regions of the $N(0,1)$ distribution (the Gaussian distribution the decoder samples from). 50,000 jet image are used to generate each curve. These curves reveal several interesting characteristics of the VAE. The mass, $p_T$, and N-subjettiness of the W boson and QCD jet images shift in various directions throughout the $N(0,1)$ distribution. A key pattern that we can see is that the spikes in the jet observable graphs start to flatten as the latent space value approaches 1. We visually inspect these patterns in Figure \ref{LIAJI}, which displays the average jet image for W boson and QCD jet images for each interval of the $N(0,1)$ distribution and the difference between the two images in the middle column. A pattern that can be observed is that the outer radiation of the QCD jet image and the prongs of the W boson jet images start to become more prominent as they approach 1.

Due to the wide variety of curves that is produced by the VAE, the VAE can be potentially useful for certain tasks such as jet tagging. Specific jet images produced by VAE can be used to test certain taggers in distinguishing signal from background generated by various regions in $N(0,1)$. For example, we can test how well a tagger can tag jet images in the [0.75, 1] interval where the mass, $p_T$, and N-subjettiness distributions are very close together. By increasing the number of intervals in $N(0,1)$ , a wider variety of curves can be seen. For this assessment, we only use 8 intervals from this Gaussian distribution but a higher number can be used. 

\begin{figure}[h!]
  \centering
  \includegraphics[width=0.7\textwidth]{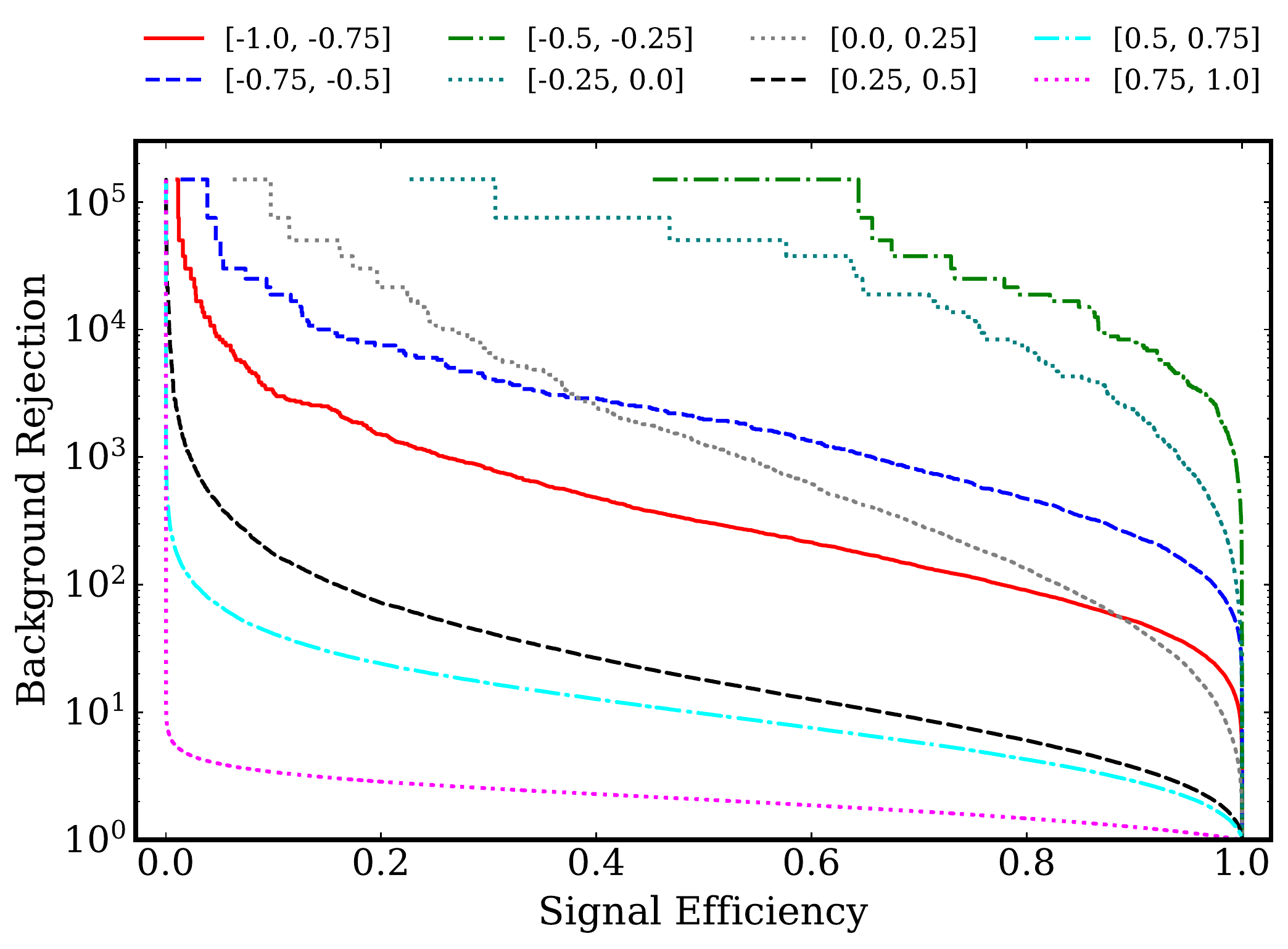}
  \caption{ROC curves (signal efficiency vs background rejection) for jet images, with a cut on mass, produced by different regions in $N(0,1)$. This illustrates how the cut between W boson and QCD jet images can be controlled by the VAE.}
  \label{ROC_mass}
\end{figure}

\begin{figure}[h!]
  \centering
  \includegraphics[width=0.7\textwidth]{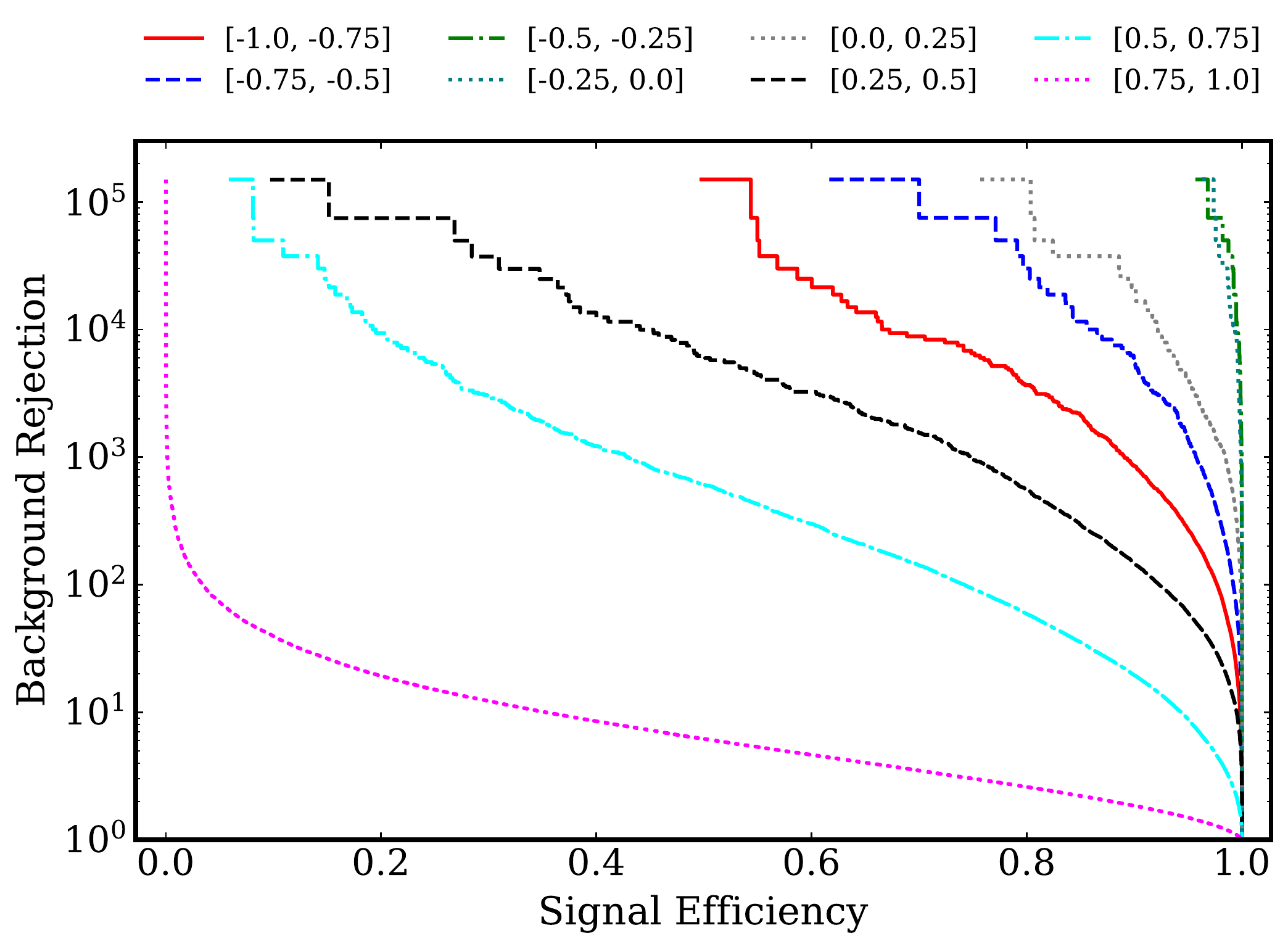}
  \caption{ROC curves (signal efficiency vs background rejection) for jet images, with a cut on $p_T$, produced by different regions in $N(0,1)$.}
  \label{ROC_pt}
\end{figure}

\begin{figure}[h!]
  \centering
  \includegraphics[width=0.9\textwidth]{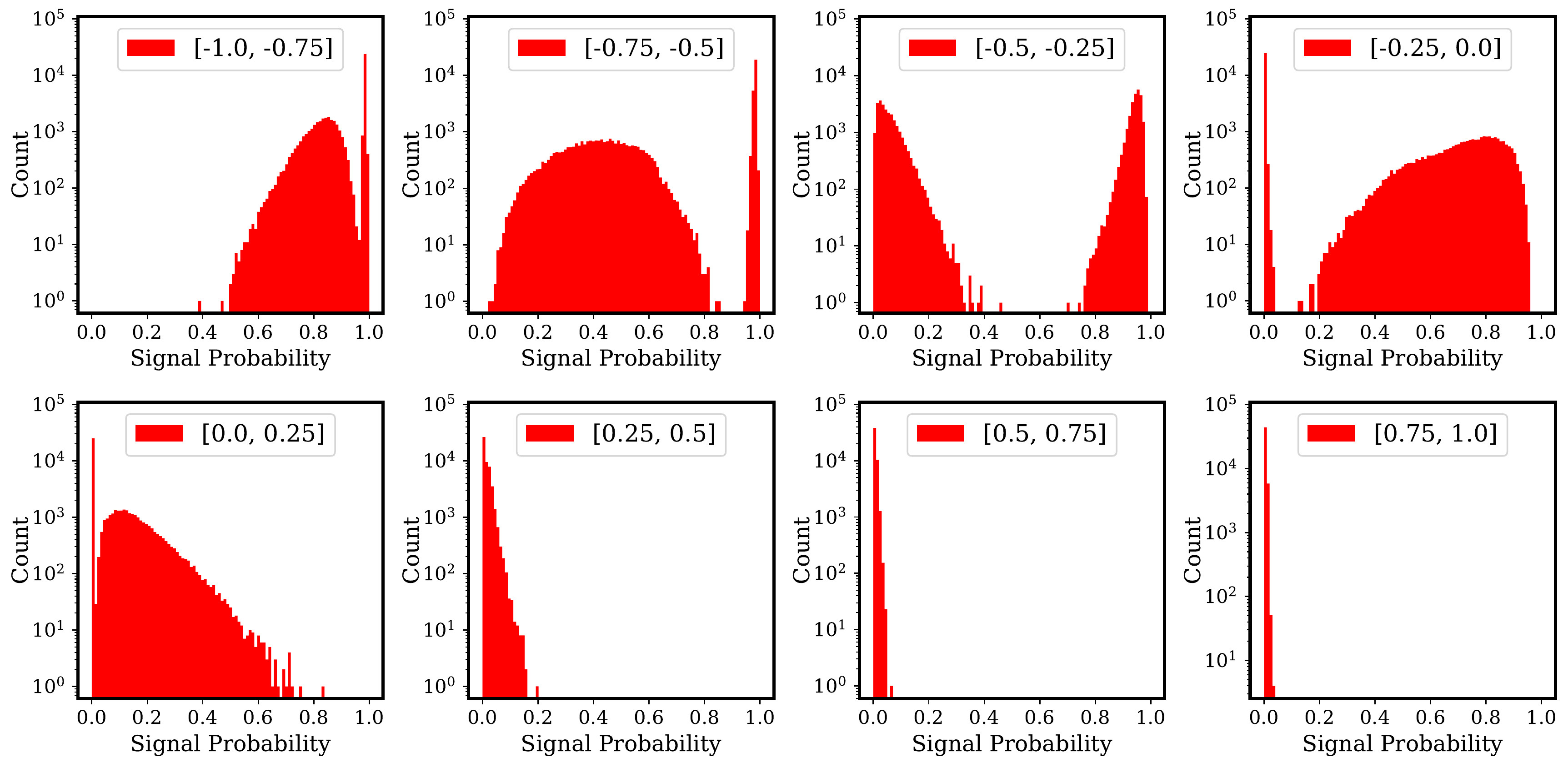}
  \caption{Classifier predictions of signal and background (0 represents background and 1 represents signal) for various regions in N(0,1).}
  \label{N_class}
\end{figure}

To support our claim, we plot the ROC curve of signal efficiency vs background rejection for various regions in N(0,1). We test this on the mass and $p_T$ of the jet images in Figures \ref{ROC_mass} and \ref{ROC_pt}. The ROC curves demonstrate how the cut between signal and background can change for different kinds of jet images produced by the VAE. As shown in Figures \ref{LIPD} and \ref{LIPT}, the distributions are harder to cut as we approach 1 in the N(0,1) distribution. This is shown by the low Area Under the Curve (AUC) in Figures \ref{ROC_mass} and \ref{ROC_pt} from [0.75-1]. We can further see how the distinctness of QCD and W boson jet images change in Figure \ref{N_class}. Because similar masses between jet images can be obtained from either two hard prongs or with one hard prong with a diffused spray, a more visual assessment can be observed by seeing how a classifier predicts W boson and QCD for various regions in N(0,1). This supports Figure \ref{LIAJI}, where although the jet images have similar mass and $p_T$ in the [0.75,1] region, they have visually different features. 


\begin{figure}[h!]
  \centering
  \includegraphics[width=0.82\textwidth]{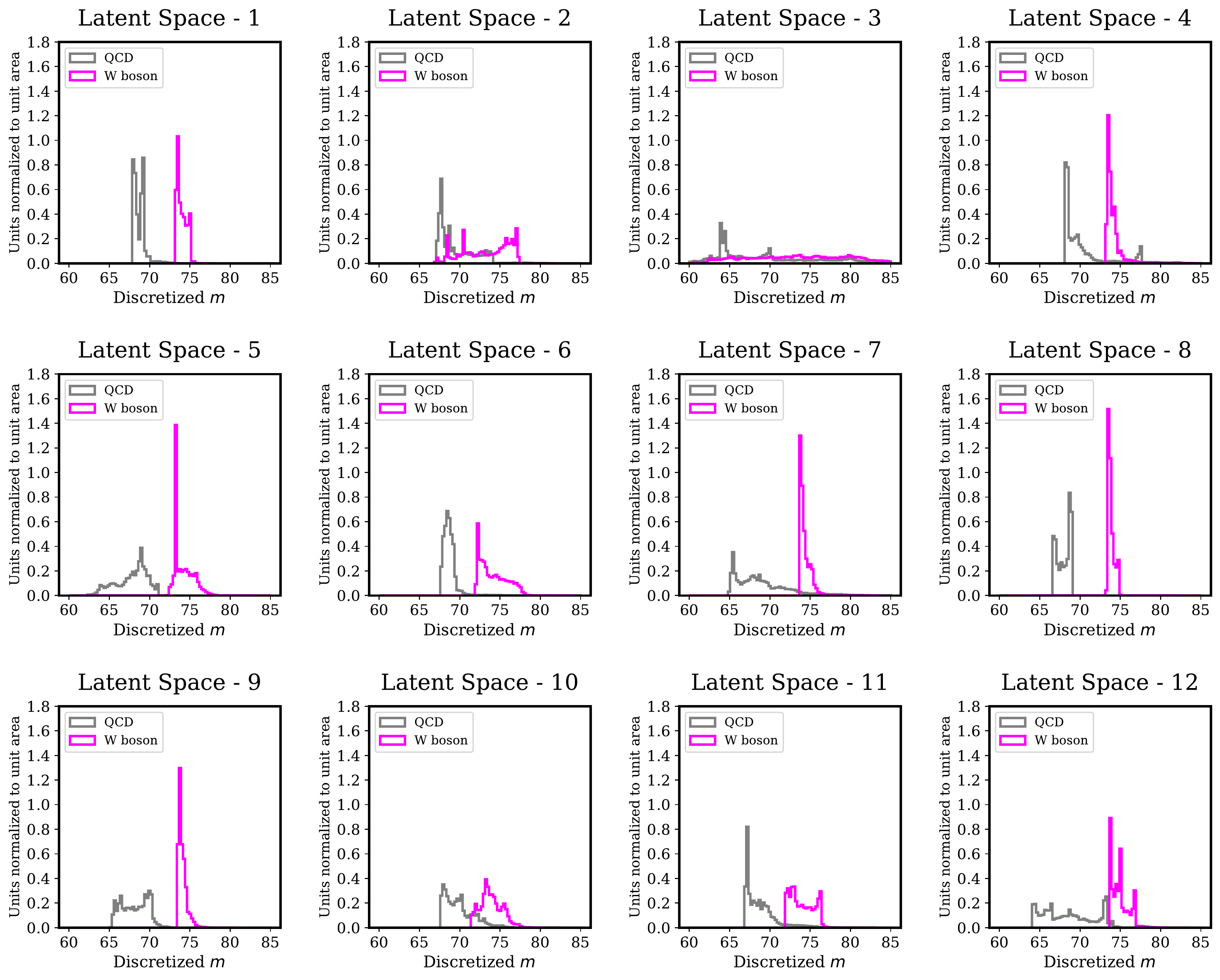}
  \caption{Mass distributions for each latent space value.}
  \label{LSVM}
\end{figure}

\begin{figure}[h!]
  \centering
  \includegraphics[width=0.82\textwidth]{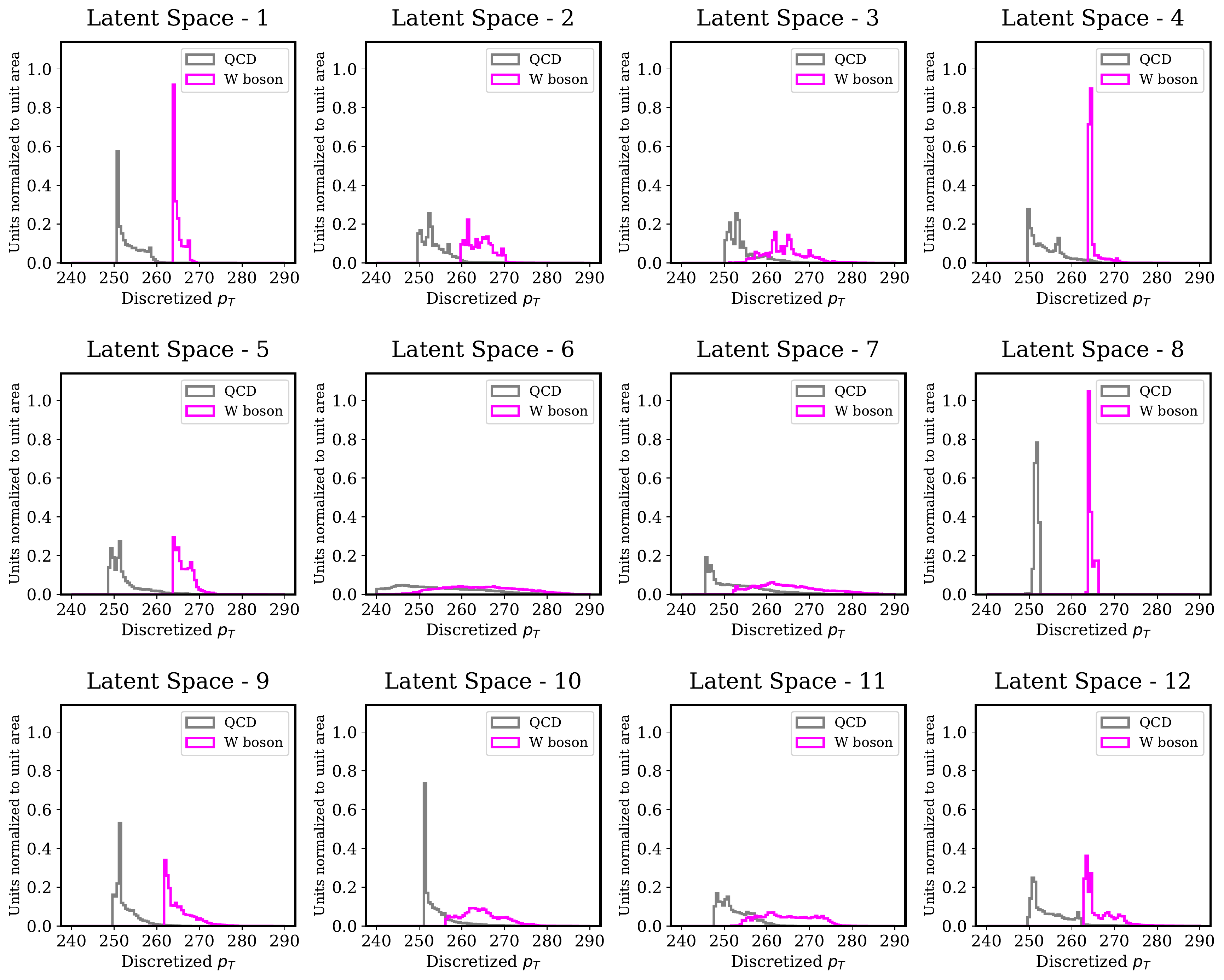}
  \caption{$p_T$ distributions for each latent space value.}
  \label{LSVPT}
\end{figure}

We further investigate the latent space by observing how each latent space value affects the overall distribution of data being produced. To analyze this, we set all other values in the latent size to zero except for the one we are analyzing and calculate the mass and $p_T$ for the generated images. These graphs are shown in Figures \ref{LSVM} and \ref{LSVPT}. A diverse range of distributions is shown for different latent space values. These graphs are not necessarily realistic distributions of data because a VAE will normally utilize all latent space values, but these figures highlight what kind of information is scattered throughout a latent space of 12. There are some subtle patterns found in the mass and $p_T$ curves but not necessarily strong ones. For example, we can see a large spike in mass for the jet images produced from latent space values of 7 through 9 and a smoothing of the distributions generated by latent space values of 2 and 3. All latent space values are ultimately used by the VAE to create an aggregate of all these features. This is why the W boson jet images have large spikes in mass towards the center of the curves because it helps create the spike in mass found in the W boson mass distribution in Figure \ref{Distributions}.

\begin{figure}[h!]
  \centering
  \includegraphics[width=1\textwidth]{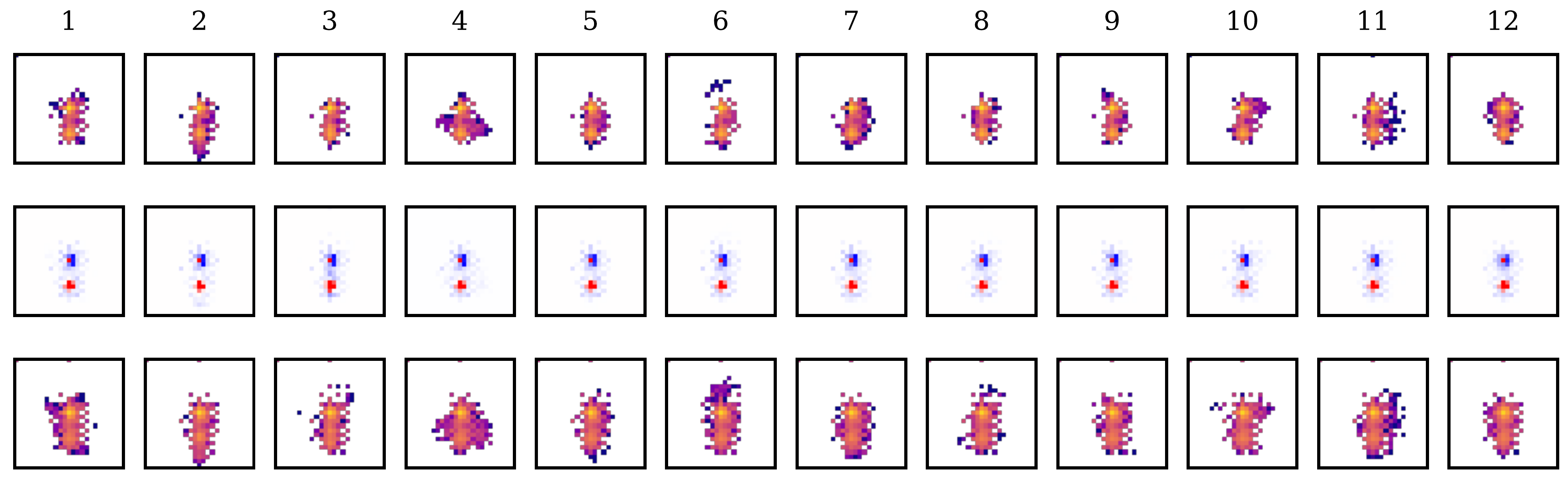}
  \caption{Average jet images for each latent space value. The top column represents the average W boson jet image, the bottom represents the average QCD jet image, and the center column is the difference between the two images. We can observe how the W boson and QCD jet images have unique properties but the difference between them remain relatively the same throughout all latent space values.}
  \label{LSAJIP}
\end{figure}

We visualize the curves in Figures \ref{LSVM} and \ref{LSVPT} by plotting the average W boson image and the average QCD image for each latent space value and the difference between both images in Figure \ref{LSAJIP}. We can see how there is little variation between the center images throughout different latent space values but the features of the jet images change. This illustrates that the latent space of the VAE produces W boson and QCD jet images with unique features but the difference between them are consistent throughout all latent space values. 

\section{Speed Comparison}
\begin{figure}[h!]
\begin{subfigure}{.5\textwidth}
  \centering
  \hspace*{0em}
  \includegraphics[width=0.75\linewidth]{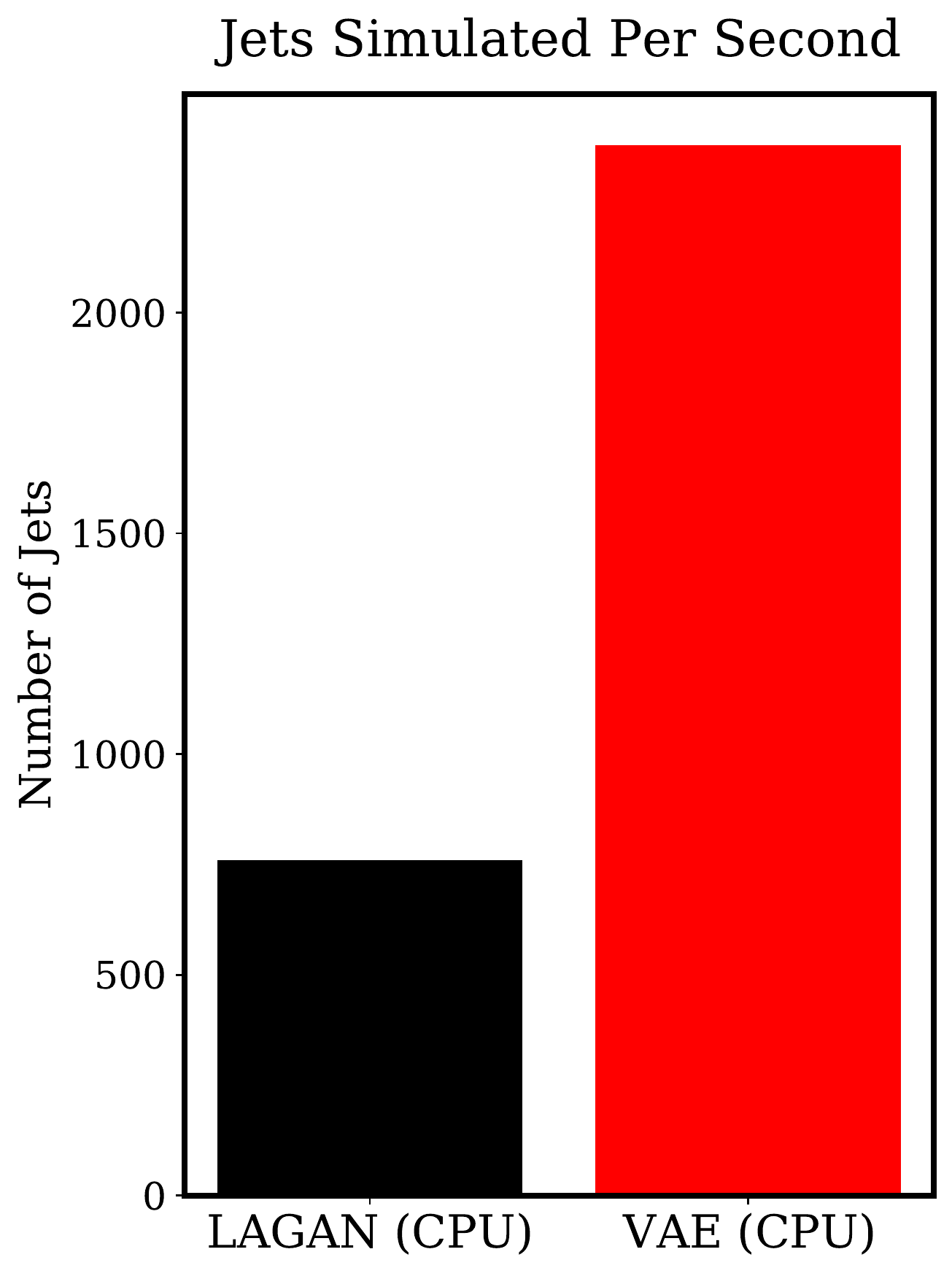}  
\end{subfigure}
\begin{subfigure}{.5\textwidth}
  \centering
  \hspace*{-5em}
  \includegraphics[width=0.75\linewidth]{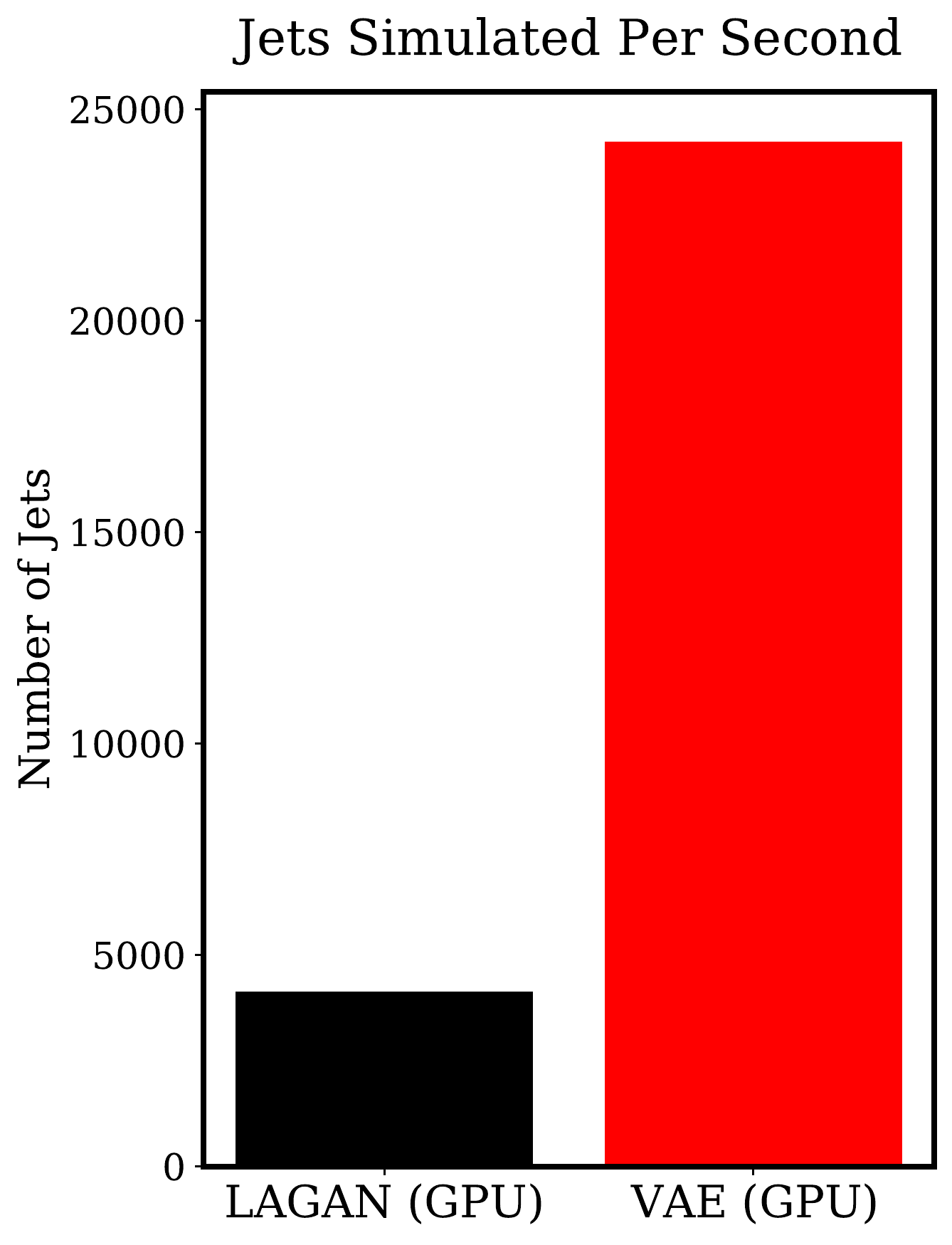}  
\end{subfigure}
\caption{Speed Comparison between LAGAN and VAE using both a CPU and a GPU.}
\label{Time Comparison}
\end{figure}

Speed is a very important factor when developing a deep generative model for high energy physics. Due to the increasing reliance on simulation at the LHC, generative models need to be not only accurate in simulating detail data, but they also have to be fast. The current most popular generative model to generate jet images is the GAN. To test the speed of the VAE, we compare the total time it takes to generate 300,000 jet images and its rate of producing jet images with the LAGAN in Table \ref{table} and Figure \ref{Time Comparison}. 

\begin{table}[h!]
\centering
\caption{Speed Comparison of LAGAN and VAE}
\begin{tabular}[t]{lcc}
\hline
&Events/s & Time for 300k Events (s)\\
\hline
LAGAN (CPU)&714&420\\
LAGAN (GPU)&4102&73\\
VAE (CPU)&2176&138\\
VAE (GPU)&24209&12\\
\hline
\end{tabular}
\label{table}
\end{table}%

The VAE model, using a GPU, is shown to be around 6 times faster than the LAGAN. This is most likely the case because the LAGAN depends on locally-connected layers to generate the radiation patterns and jet prongs of its jet images. Locally-connected layers are computationally expensive, therefore, it would make sense that the LAGAN will take up more time to generate its jet images. This contrasts our model, which uses convolutional layers to obtain the distinct features of jet images. 

In addition to the simulation speed of VAEs, the VAE is also shown to be easy to train. Because the structure of the VAE does not require adversarial training, which is known to be difficult to train, it can optimize relatively more smoothly. This ultimately encourages the exploration of hybrids of GANs and VAEs. 

\section{Conclusion}

In this paper, we have successfully implemented a variational autoencoder to simulate realistic jet images. The model we propose is able to reproduce key characteristics of Pythia jet images. Some of these characteristics include mass, $p_T$, N-subjettiness, and pixel intensity. In addition to showing the accuracy of the VAE, we explored the latent space and saw how it can be used to produce unique distributions. This can be potentially used for other tasks such as jet tagging, where we can assess how certain taggers perform when given a wide variety of jet images produced by the VAE. The VAE we propose is also demonstrated to be faster than the LAGAN due to its lack of reliance on computationally expensive architectures such as locally-connected layers. 

This paper also demonstrated the effectiveness of the feature perceptual loss for computer-vision-related techniques in high energy physics. This can be applied to not only other types of generative models, but they can also be applied to other algorithms for specific tasks such as autoencoders for anomaly detection. The feature perceptual can replace any standard distance metric loss such as the mean squared error, making it a powerful, yet simple technique for high energy physics tasks that require image recognition.

GANs have been known to produce images that are sharper than VAEs. We explore new and well-established VAE techniques in machine learning literature to ultimately create a fast algorithm that is comparable to the GAN in performance. Because the GAN and VAE have their own unique advantages, exploring hybrid models seem to be a promising direction for future research. To summarize, the VAE we propose is ultimately a fast, easy-to-train, and accurate generative model that can be used to generate realistic high energy physics data.

\section*{Acknowledgments}

I would like to thank my mentor Dr. David Shih for guiding this research. He has provided useful discussions, background material on machine learning in high energy physics, feedback on the manuscript, and overall great support. I would also like to thank Dr. Benjamin Nachman for his valuable feedback on the manuscript and for insightful discussion. 

{
  \hypersetup{urlcolor=blue}
  \printbibliography
}

\end{document}